\documentclass[%
reprint,
amsmath,amssymb,
aps,
]{revtex4-1}
\usepackage{mathrsfs}
\usepackage{graphicx}
\usepackage{amsfonts}
\usepackage{graphicx}
\usepackage{dcolumn}
\usepackage{bm}
\usepackage[colorlinks,urlcolor=blue,linkcolor=blue,citecolor=blue]{hyperref}

\begin{document}

\title{Multiple-photon bundle emission in the $n$-photon Jaynes-Cummings model}
\author{Shu-Yuan Jiang}
\email{These authors contributed equally to this work.}
\affiliation{Key Laboratory of Low-Dimensional Quantum Structures and Quantum Control of
Ministry of Education, Key Laboratory for Matter Microstructure and Function of Hunan Province, Department of Physics and Synergetic Innovation Center for Quantum Effects and Applications, Hunan Normal University, Changsha 410081, China}

\author{Fen Zou}
\email{These authors contributed equally to this work.}
\affiliation{Key Laboratory of Low-Dimensional Quantum Structures and Quantum Control of
Ministry of Education, Key Laboratory for Matter Microstructure and Function of Hunan Province, Department of Physics and Synergetic Innovation Center for Quantum Effects and Applications, Hunan Normal University, Changsha 410081, China}

\author{Yi Wang}
\affiliation{Key Laboratory of Low-Dimensional Quantum Structures and Quantum Control of
	Ministry of Education, Key Laboratory for Matter Microstructure and Function of Hunan Province, Department of Physics and Synergetic Innovation Center for Quantum Effects and Applications, Hunan Normal University, Changsha 410081, China}

\author{Jin-Feng Huang}
\email{Corresponding author: jfhuang@hunnu.edu.cn}
\affiliation{Key Laboratory of Low-Dimensional Quantum Structures and Quantum Control of
Ministry of Education, Key Laboratory for Matter Microstructure and Function of Hunan Province, Department of Physics and Synergetic Innovation Center for Quantum Effects and Applications, Hunan Normal University, Changsha 410081, China}

\author{Xun-Wei Xu}
\email{Corresponding author: xwxu@hunnu.edu.cn}
\affiliation{Key Laboratory of Low-Dimensional Quantum Structures and Quantum Control of
Ministry of Education, Key Laboratory for Matter Microstructure and Function of Hunan Province, Department of Physics and Synergetic Innovation Center for Quantum Effects and Applications, Hunan Normal University, Changsha 410081, China}

\author{Jie-Qiao Liao}
\email{Corresponding author: jqliao@hunnu.edu.cn}
\affiliation{Key Laboratory of Low-Dimensional Quantum Structures and Quantum Control of
Ministry of Education, Key Laboratory for Matter Microstructure and Function of Hunan Province, Department of Physics and Synergetic Innovation Center for Quantum Effects and Applications, Hunan Normal University, Changsha 410081, China}

\begin{abstract}
We study the multiple-photon bundle emission in the $n$-photon Jaynes-Cummings model composed of a two-level system coupled to a single-mode optical field via the $n$-photon exciting process. Here, the two-level system is strongly driven by a near-resonant monochromatic field, and hence the system can work in the Mollow regime, in which a super-Rabi oscillation between the zero-photon state and the $n$-photon state can take place under proper resonant conditions. We calculate the photon number populations and the standard equal-time high-order correlation functions, and find that the multiple-photon bundle emission can occur in this system. The multiple-photon bundle emission is also confirmed by investigating the quantum trajectories of the state populations and both the standard and generalized time-delay second-order correlation functions for multiple-photon bundle. Our work paves the way towards the study of multiple-photon quantum coherent devices, with potential application in quantum information sciences and technologies.
\end{abstract}
\maketitle


\section{Introduction}

Multiple-photon bundle emission~\cite{10.1038/nphoton.2014.114}, as one of the physical mechanisms for preparation of multiple-photon sources, has recently attracted much attention from researchers in various fields of physics, such as quantum optics, quantum information, and laser physics. This is because multiple-photon sources not only are very useful to the study of fundamental quantum physics, but also have wide application potential in quantum information sciences, including quantum communication \cite{Kimble2008Jun}, quantum lithography \cite{PhysRevLett2000,Two-photon2001}, quantum cryptography~\cite{Gisin2002Mar}, and quantum metrology \cite{Giovannetti1330,PhysRevLett.96.010401}. Until now, various methods for generation of multiple-photon states have been proposed in atom-coupled photonic waveguides~\cite{ PhysRevX.6.031017,PhysRevLett.118.213601,Chang2018Aug}, Rydberg
atomic ensembles~\cite{PhysRevA.90.053804,PhysRevLett.115.123601}, Kerr cavity systems~\cite{Liao2010Nov}, cavity optomechanical systems~\cite{liao2013Correlated,Qin2019Dec}, cavity quantum electrodynamics (QED) systems~\cite{Haroche2006Oct}, and multiple-level atomic systems~\cite{PhysRevLett.117.203602,PhysRevA.100.043825,Muller2014Mar,Hargart2016Mar,Koshino2013Oct,Dousse2010Jul,Ota2011Nov,Sanchez-Burillo2016Nov}. Unlike the photon blockade effect~\cite{xn--Imamolu-4s3c1997Aug,Birnbaum2005Jul,Liew2010May,Rabl2011Aug,Ridolfo2012Nov,Liao2013Aug,Miranowicz2013Feb,Snijders2018Jul,Vaneph2018Jul,Huang2018Oct,Zou2019Apr,Xu2020Feb,Zou2020Nov,Ren2021May,Deng2021Sep}, the energy unit of the multiple-photon bundle emission is a bundle of several photons rather than a single photon. In addition, the physical mechanism for multiple-photon bundle emission is a bundle of photons blocking the transmission of the next bundle of photons, rather than $n$ photons blocking the transmission of the ($n+1$)th photon. For the sake of application, the implementation of a controllable multiple-photon source, namely a multiple-photon gun, is a desired task in quantum information. Therefore, the multiple-photon bundle emission becomes a significant research topic, because it provides a physical mechanism for the implemention of a multiple-photon gun.

Recently, much effort has been devoted to the study of multiple-photon (-phonon) bundle emission. A range of schemes for $N$-photon (phonon) bundle emission have been proposed in various quantum systems, e.g., cavity-QED systems~\cite{Strekalov2014Jul, 10.1364/OPTICA.5.000014,PhysRevLett.124.053601,bin2020paritysymmetryprotected,Deng:21,Cosacchi2021Aug,Diaz-Camacho2021Sep}, circuit-QED systems~\cite{Ma2021Oct}, and cavity optomechanical systems~\cite{Zou2021Dec}. In particular, the  bundle emission of photons has been observed in a dc-biased superconducting circuit~\cite{Menard2022Apr}. In general, to achieve $N$-photon bundle emission, it is needed to generate the $N$-photon state in advance. Currently, several mechanisms for geneartion of multiple-photon states have been proposed, such as the photon-number state climbing process~\cite{Hofheinz2008Jul} and high-order process of the Jaynes-Cummings (JC) coupling~\cite{10.1038/nphoton.2014.114}. We point out that for these two mentioned mechanisms, the $n$th-order processes of single-photon coupling are needed to create an $n$-photon state. Here, the order of a physical process is determined by the transition matrix elements. According to the theory for perturbation calculation of transition amplitudes, the order of the physical process associated with the transition matrix element $\langle\psi_{f}\vert H^{n}\vert\psi_{i}\rangle$ is $n$, where $H$ is the Hamiltonian of the interaction, and $\vert\psi_{f}\rangle$ and $\vert\psi_{i}\rangle$ are the final state and initial state, respectively. In general, compared with \textit{the high-order single-photon physical process}, \textit{the first-order multiple-photon process} will lead to a higher probability for generation of the $n$-photon states under the same coupling strength in the weak-coupling regime. Therefore, a natural question is whether one can use the first-order multiple-photon process to generate the multiple-photon states. Here, the first-order multiple-photon process means that the multiple photons are simultaneously involved in the physical process, rather than involving multiple steps. Note that the $n$-photon process could also be induced by an $n$th-order perturbation of single-photon process.

Motivated by this point, in this work we present a scheme for implementing multiple-photon bundle emission in the $n$-photon JC model consisting of a two-level system (TLS) and a single-mode optical field. Here, the TLS is coupled to the optical field via the $n$-photon JC interaction, and the TLS is driven by a monochromatic field. When the driving is much stronger than the $n$-photon JC coupling, the $n$-photon JC interaction can be taken as a perturbative term. By analyzing the energy spectrum of the system in the Mollow regime~\cite{Mollow1969Dec,Schuda1974May,Kimble1976Jun,Jetter238U,Gonzalez_Tudela_2013,article}, we find that the resonant oscillation between the zero-photon state and the $n$-photon state can be realized via the $n$-photon JC interaction.
Here, the Mollow regime indicates that the TLS is driven by a strong laser near resonance. We confirm this super-Rabi oscillation by analytically and numerically calculating the state populations. To investigate the multiple-photon bundle emission, we analyze the photon number populations, the standard $\ell $th-order ($\ell =2, 3,~\textrm{and}~4$) correlation functions, and the quantum trajectory of the state populations by numerically solving the quantum master equation. We also calculate the standard and generalized time-delayed second-order correlation functions for multiple-photon bundle to characterize the quantum statistical properties of the optical mode. The results indicate that the $n$-photon JC system can behave as a multiple-photon gun by choosing appropriate resonant transitions. This scheme will have wide applications in quantum information processing and other tasks based on multiple-photon sources.

\section{Model and Hamiltonian}\label{sectionII}

\begin{figure}
\begin{centering}
\includegraphics[scale=0.33]{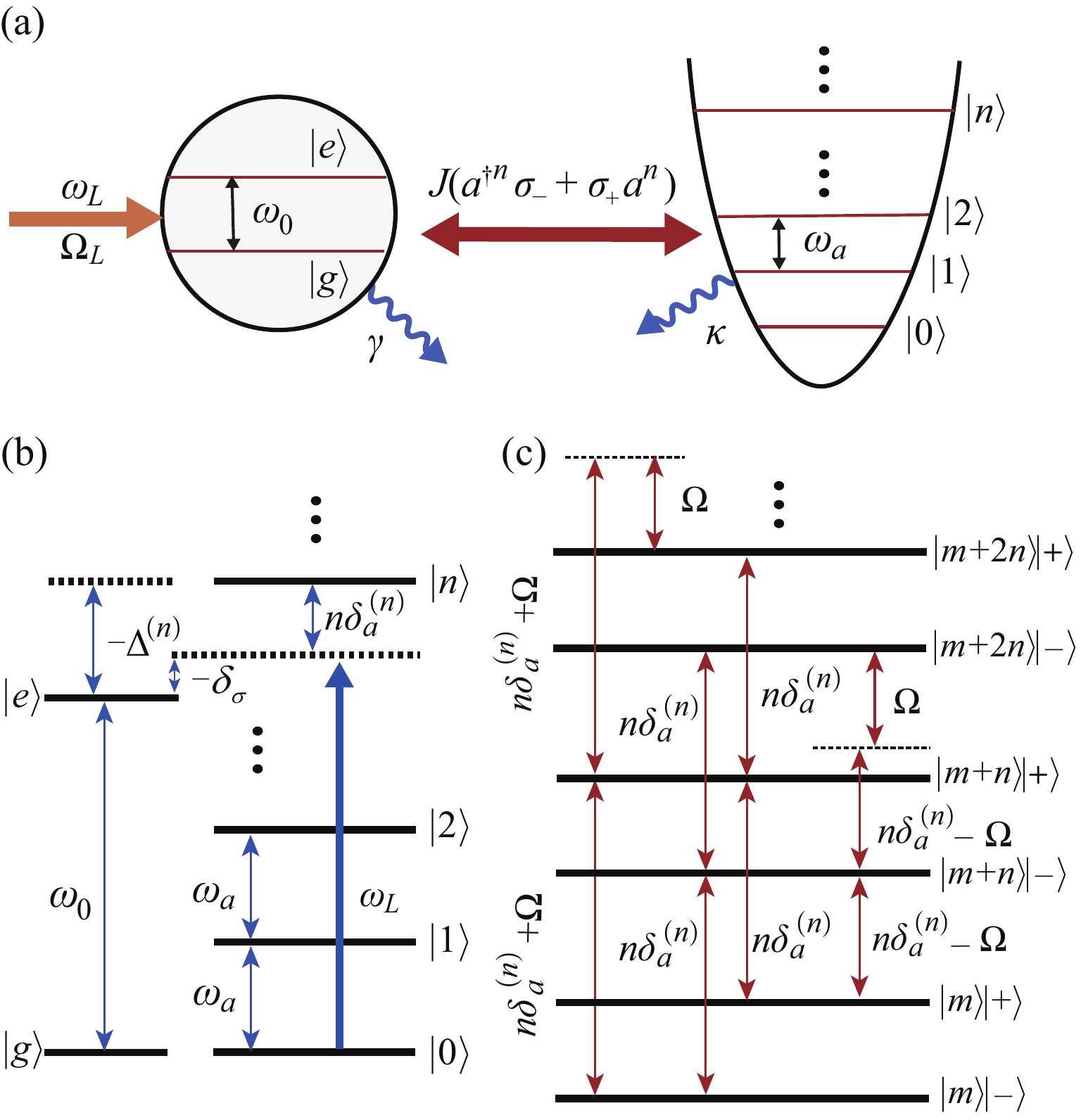}
\par\end{centering}
\centering{}\caption{(a)
Schematic of the $n$-photon JC model composed of a single optical mode coupled to a TLS via the $n$-photon JC process. (b)
Energy-level diagram of the bare states of the system in the Schr\"{o}dinger picture. (c) Energy levels of the system in the Mollow regime. Here, the energy levels are expressed in the eigen-representation of the Hamiltonian $H_{0}$ in Eq.~(\ref{H_{0}}), and the transitions occur among the $m$-,~$(m+n)$-, and $(m+2n)$-photon domains.}\label{Fig1}
\end{figure}

We consider the $n$-photon JC model~\cite{SUKUMAR1981211,PhysRevA.25.3206,Villas-Boas2019Mar,Larson_2021}, which describes the interaction of a TLS with a single-mode optical field via the $n$-photon exciting process, as shown in Fig.~{\ref{Fig1}}(a). Note that here we use the terms of the TLS and optical mode to describe the system. However, the $n$-photon JC model described in the work is a general physical model, and hence the present scheme can be implemented in any physical platforms with which the $n$-photon JC model can be realized. Since the excitations of a bosonic mode could be either phonons or photons,
the physical mechanism proposed in this work can also be used to implement multiple-phonon bundle emission. For keeping representation clear, we will use the terms of TLS and photons throughout this work.

For the $n$-photon JC model~\cite{SUKUMAR1981211,PhysRevA.25.3206,Villas-Boas2019Mar,Larson_2021}, it is described by the Hamiltonian ($\hbar=1$)
\begin{equation}
H_{\textrm{npJC}}=\omega_{a}a^{\dagger}a+\omega_{0}\sigma_{+}\sigma_{-}+J(a^{\dagger n}\sigma_{-}+\sigma_{+}a^{n}), \label{1}
\end{equation}
where $a~(a^{\dagger})$ is the annihilation (creation) operator of the optical mode with resonance  frequency $\omega_{a}$, and the operator $\sigma_{-}=|g\rangle \langle e|~(\sigma_{+}=|e\rangle\langle g|)$
is the lowering (raising) operator of the TLS with the transition frequency $\omega_{0}$ between the excited state $|e\rangle $ and the ground state $|g\rangle $. The parameter $J$
is the coupling strength of the $n$-photon JC interaction, and $n$ is the photon number associated with each transition process of the TLS. We consider the case where the TLS is continuously  driven by a laser with frequency $\omega_{L}$ and amplitude $\Omega_{L}$. The driving Hamiltonian is described by
\begin{eqnarray}
 H_{L}&=&\Omega_{L}(\sigma_{+}e^{-i\omega_{L}t}+\sigma_{-}e^{i\omega_{L}t}).\label{2}
\end{eqnarray}
In a rotating frame defined by the
unitary operator
$\textrm{exp}\{-(i\omega_{L}t/n)[a^{\dagger}a+(n/2)\ensuremath{\sigma_{z}}]\}$,
the total Hamiltonian of the system becomes
\begin{equation}
H_{I}=\delta_{a}^{(n)}a^{\dagger}a+\delta_{\sigma}\sigma_{+}\sigma_{-}+J(a^{\dagger n}\sigma_{-}+\sigma_{+}a^{n})+\Omega_{L}\sigma_{x},\label{HI}
\end{equation}
where $\delta_{a}^{(n)}=\omega_{a}-\omega_{L}/n$ is the single-cavity-photon process detuning, and $\delta_{\sigma}=\omega_{0}-\omega_{L}$
is the atomic driving detuning, as shown in Fig.~\ref{Fig1}(b). Note that a constant term $\omega_{L}/2$ has been omitted in Eq.~(\ref{HI}).

\section{Multiple-photon bundle emission in the Mollow regime}\label{sectionIII}

In this section, we analyze the eigensystem of the system in the Mollow regime, and discuss the super-Rabi oscillation between the zero-photon state and the $n$-photon ($n=2$ and $3$) state. We also study the multiple-photon bundle emission by examining the photon number populations, the standard equal-time high-order correlation functions, the Monte Carlo simulations of state populations, and both the standard and generalized time-delay second-order correlation functions for multiple-photon bundle.

\subsection{Super-Rabi oscillation}\label{subsectionA}

The super-Rabi oscillation provides a clear physical mechanism for creating the $n$-photon state. It has been shown that for the JC model, both the Mollow regime and JC-coupling regime can be used to implement the super-Rabi oscillation~\cite{10.1038/nphoton.2014.114}. In the present model, we find that the Mollow regime is better to be used to realize the bundle emission. As a result, below we mainly focus on the Mollow regime, and present some discussions concerning the $n$-photon JC coupling regime in Sec.~\ref{sectionIV}. In the Mollow regime, the $n$-photon JC couping strength $J$ is much smaller than the driving amplitude $\Omega_{L}$, and then the $n$-photon JC coupling term can be treated as a perturbation. Up to the zero order of the JC coupling, the Hamiltonian related to the TLS becomes
\begin{equation}
H_{\sigma}=\delta_{\sigma}\sigma_{+}\sigma_{-}+\Omega_{L}\sigma_{x}.
\end{equation}
The eigenvalues and eigenstates of the Hamiltonian $H_{\sigma}$
are given by $E_{\pm}=({\delta_{\sigma}}\pm\Omega)/2$ and $\vert \pm \rangle=c_{\pm}\vert e\rangle \pm c_{\mp}\vert g\rangle$. Here the superposition coefficients are given by $c_{\pm}=\sqrt{2\Omega_{L}^{2}/(\Omega^{2}\mp\delta_{\sigma}\Omega)}$ satisfying the normalization condition $c_{+}^{2}+c_{-}^{2}=1$, where we introduced the generalized Rabi frequency $\Omega=\sqrt{\delta_{\sigma}^{2}+4\Omega_{L}^{2}}$. To study the physical processes induced by the $n$-photon JC interaction, below we work in a rotating frame with respect to the Hamiltonian
\begin{equation}
H_{0}=\delta_{a}^{(n)}a^{\dagger}a+\delta_{\sigma}\sigma_{+}\sigma_{-}+\Omega_{L}\sigma_{x},\label{H_{0}}
\end{equation}
which has the eigensystem
\begin{equation}
H_{0}\vert m\rangle \vert \pm \rangle
=(E_{\pm }+m\delta_{a}^{(n)}) \vert m\rangle
\vert \pm \rangle .
\end{equation}
In this frame, the Hamiltonion $H_{I}$ becomes
\begin{align}
V_{I}(t)=\sum_{m=0}^{\infty }\sum_{s,r=\pm }A_{s,r}B_{s,r,m}(t)
|m+n\rangle |s\rangle \langle
 m|\langle r|+\textrm{H.c.},\label{VI}
\end{align}
where we introduce
\begin{subequations}
\begin{align}
B_{s,r,m}(t)&=J\sqrt{\frac{(m+n)!}{m!}}e^{i(
E_{s}-E_{r}+n\delta _{a}^{(n)})t },\\
A_{+,+}&=\langle +\vert \sigma _{-}\vert +\rangle
=c_{+}c_{-},\\ A_{+,-}&=\langle
+\vert \sigma _{-}\vert -\rangle =c_{-} ^{2},\\
A_{-,+}&=\langle -\vert \sigma _{-}\vert +\rangle
=-c_{+}^{2} ,\\
A_{-,-}&=\langle -\vert
\sigma _{-}\vert -\rangle =-c_{+}c_{-}.
\end{align}
\end{subequations}

It can be seen from Eq.~(\ref{VI}) that the $n$-photon JC coupling will induce the state transitions between the $m$-photon domain $(|m\rangle|r\rangle~\textrm{for}~r=\pm)$ and the $(m+n)$-photon domain $(|m+n \rangle|s\rangle~\textrm{for}~s=\pm)$, accompanied by a change of $n$ photons. We point out that the merit for the introduction of the $n$-photon JC coupling is the increasing of $m$ photons by the first-order process. This point is completely different from the JC-coupling case, in which the $n$th-order process are needed to increase $n$ photons. This difference indicates the merit of the first-order $n$-photon JC interaction and brings out the motivation of this work. In particular, for the JC model, if we consider proper parameter conditions to effectively obtain the common $n$-photon coupling term, there will inevitably induce other additional coupling terms, and these additional terms will cause various physical processes, which will affect the bundle emission process. This inspires us to study the bundle emission based on the pure $n$-photon JC interaction. In this case, there is no influence caused by those additional coupling terms, and then we can clearly know the physical effect and parameter condition for the multiple-photon bundle emission.

We can see from Eq.~(\ref{VI}) that, for the first-order physical process associated with the $n$-photon JC Hamiltonian, there are four processes for the transitions between the $m$- and $(m+n)$-photon domains: $|m\rangle|\pm\rangle\leftrightarrow |m+n\rangle|\pm\rangle$. The transition frenquencies corresponding to $|m\rangle| +\rangle\leftrightarrow|m+n\rangle|-\rangle$ and $|m\rangle|-\rangle\leftrightarrow|m+n\rangle|+\rangle$ are $n\delta_{a}^{(n)}-\Omega$ and $n\delta_{a}^{(n)}+\Omega$, respectively. In addition, the transition frequencies corresponding to both the two transitions $|m\rangle|+\rangle\leftrightarrow|m+n\rangle|+\rangle$ and $|m\rangle|-\rangle\leftrightarrow|m+n\rangle|-\rangle$ are $n\delta_{a}^{(n)}$. To realize a perfect super-Rabi oscillation, we should choose proper initial state and resonant condition such that the Hilbert space can be approximately truncated. To this end, we should choose either the transition
$|m\rangle|+\rangle\leftrightarrow|m+n\rangle|-\rangle$ or the transition $|m\rangle|-\rangle\leftrightarrow|m+n\rangle|+\rangle$. This is because when the transition $|m\rangle|+\rangle\leftrightarrow|m+n\rangle|-\rangle$ is resonant, then the transitions $|m+n\rangle|-\rangle\leftrightarrow|m+2n\rangle|+\rangle$ and $|m+n\rangle|-\rangle\leftrightarrow|m+2n\rangle|-\rangle$ will be supressed by the detunings $2\Omega$ and $\Omega$, respectively. Similiarly, when the transition $|m\rangle|-\rangle\leftrightarrow|m+n\rangle|+\rangle$ is resonant, then the transitions $|m+n\rangle|+\rangle\leftrightarrow|m+2n\rangle|-\rangle$ and $|m+n\rangle|+\rangle\leftrightarrow|m+2n\rangle|+\rangle$ will be detuned by $2\Omega$ and $\Omega$, respectively. Corresponding to the above two cases, the system can be approximately restricted into the two subspaces with the bases $\{|m\rangle|+\rangle, |m+n\rangle|-\rangle\}$ and $\{|m\rangle|-\rangle, |m+n\rangle|+\rangle\}$ in the Mollow regime~$\Omega_{L}\gg J$, respectively.

Based on the energy levels in Fig.~\ref{Fig1}(c), we know that the resonant conditions associated with the transitions $|0\rangle|+\rangle\leftrightarrow|n\rangle|-\rangle$ and $|0\rangle|-\rangle\leftrightarrow|n\rangle|+\rangle$ are given by $n\delta_{a}^{(n)}-\Omega=0$ and $n\delta_{a}^{(n)}+\Omega=0$, respectively. In term of the relations $\delta_{\sigma}=\Delta^{(n)}+n\delta_{a}^{(n)}$ and $\Delta^{(n)}=\omega_{0}-n\omega_{a}$, the detuning $\delta_{a}^{(n)}$ determined by the two resonance conditions  $n\delta_{a}^{(n)}\pm\Omega=0$ has the same solution
 \begin{equation}\label{eq:9}
 \delta_{a}^{(n)}=
 -\frac{(\Delta^{(n)})^{2}+4\Omega_{L}^{2}}{2n\Delta^{(n)}},
 \end{equation}
 where $\Omega_{L}$ is the driving amplitude.

For the higher-order transitions $|0\rangle|+\rangle\leftrightarrow|\mu n\rangle|-\rangle$ and $|0\rangle|-\rangle\leftrightarrow|\mu n\rangle|+\rangle$ ($\mu \ge 2$ and $\mu$ is an integer), the resonance conditions are given by $\mu n\delta_{a,\mu}^{(n)}-\Omega=0$ and $\mu n\delta_{a,\mu}^{(n)}+\Omega=0$, respectively. In these two cases, the values of the detuning $\delta_{a,\mu}^{(n)}$ can be obtained as
\begin{equation}
\delta_{a,\mu}^{(n)}=
\frac{\Delta^{(n)}\pm\sqrt{\mu^{2}(\Delta^{(n)})^{2}+4(\mu^{2}-1)\Omega_{L}^{2}}}{n(\mu^{2}-1)},\label{deltaapm}
\end{equation}
where the signs ``$+$" and ``$-$" correspond to the former and latter cases, respectively. In our following discussions, we will consider the initial state  $|0\rangle|+\rangle$ of the system and choose the resonant transition $|0\rangle|+\rangle\leftrightarrow|n\rangle|-\rangle$, then the system can be approximately restricted into the subspace with the two basis states $\{|0\rangle|+\rangle,|n\rangle|-\rangle\}$, and the super-Rabi oscillation can occur between the two states. Below, we will consider the cases of $n=2$ and 3 for simulations. In both cases, we adiabatically eliminate the intermediate uncorrelated states and only keep the coupling between the two states, then the frequency of the super-Rabi oscillation can be approximately obtained as~\cite{10.1038/nphoton.2014.114}
\begin{equation} \Omega_\textrm{eff}^{(n)}=\frac{\sqrt{n!}Jc_{+}^{2}(n\delta_{a}^{(n)}+E_{+})E_{-}}{n!J^{2}c_{-}^{4}-(n\delta_{a}^{(n)}+E_{+})E_{-}}, \label{Omega}
\end{equation}
where these variables have been defined before.
\begin{figure}
\center
\includegraphics[width=0.47 \textwidth]{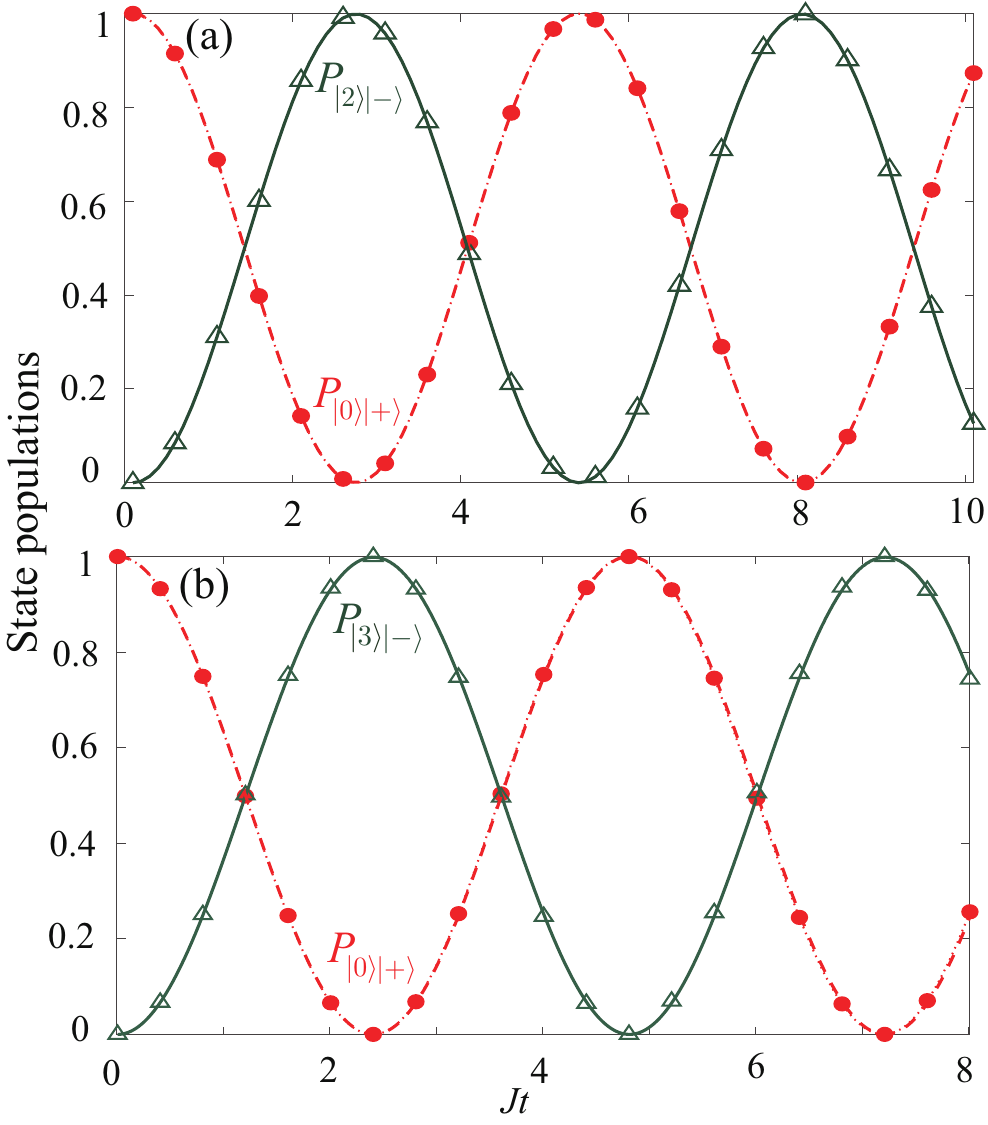}
\caption{The state populations $P_{\vert0\rangle\vert+\rangle}$ and $P_{\vert n\rangle\vert-\rangle}$ ($n=2$ and 3) as functions of the scaled evolution time $Jt$ for (a) the two-photon JC model ($n=2$) at $\Delta^{(2)}/J=-165$ and $\Omega_{L}/J=70$ and (b) the three-photon JC model ($n=3$) at $\Delta^{(3)}/J=-265$ and $\Omega_{L}/J=80$. The red dash-dotted curves and green solid curves correspond to the numerical results of the state populations, while the green triangles and the red points correspond to the analytical results based on the effective Rabi frequencies $\Omega_{\text{eff}}^{(2)}$ and $\Omega_{\text{eff}}^{(3)}$. The detuning is taken as $\delta_{a}^{(n)}\approx-{[(\Delta^{(n)})^{2}+4\Omega_{L}^{2}}]/({2n\Delta^{(n)}})$.}
\label{Fig2}
\end{figure}

In Fig.~\ref{Fig2}, we plot the populations of the states $|0\rangle|+\rangle$ and $|n\rangle|-\rangle~(n=2~\textrm{and}~3)$ as functions of the scaled evolution time $Jt$. Here, the solid and
dash-dotted lines are obtained by numerically solving the Schr\"{o}dinger equation with the full Hamiltonian in Eq.~(\ref{HI}) and the initial state $|0\rangle|+\rangle$. The markers are based on the analytical result give in Eq.~(\ref{Omega}), which is obtained by adiabatically eliminating the detuned transitions in a truncated subspace. We can see from Fig.~\ref{Fig2} that super-Rabi oscillation occur between the two states $|0\rangle|+\rangle$ and $|n\rangle|-\rangle$~for~$n=2$ and $3$. In addition, the analytical results match the numerical results well.

\subsection{Multiple-photon  bundle emission}\label{subsectionB}

In the above subsection, we have analyzed the super-Rabi oscillation between the states $|0\rangle|+\rangle$ and $|n\rangle|-\rangle$ for $n=2$ and $3$. The oscillation provides a mechanism for the preparation of the two- (three-) photon state. In the presence of the optical dissipation, the photons will be emitted and then the multiple-photon bundle emission takes place in this system. To completely describe the bundle emission process, we adopt the method of quantum master equation to govern the evolution of the system and then study the quantum statistics of the system. In the weak- and strong-coupling regimes, we can assume that the optical mode and the TSL are connected with two individual heat baths. Including the dissipations of the TLS and the optical mode, the evolution of the system is governed by the quantum master equation~\cite{scully_zubairy_1997}
 \begin{equation}
 \dot{\rho}=-i[H_{I},\rho]+\kappa\mathcal{L}[a]\rho+\gamma\mathcal{L}[\sigma_{-}]+\gamma_{\phi}\mathcal{L}[\sigma_{+}\sigma_{-}]\rho,\label{eq:21}
 \end{equation}
where the Hamiltonian $H_{I}$ is given in Eq.~({\ref{HI}}), $\kappa~(\gamma)$ is the decay rate of the optical mode (TLS), and $\gamma_{\phi}$ is the pure dephasig rate of the TLS. The standard Lindblad super-operators are defined by $\mathcal{L}[o]\rho=(2o\rho o^{\dagger}-\rho o^{\dagger}o-o^{\dagger}o\rho)/2$ for $o=a$ and $\sigma_{-}$.  For the case of multiple-photon bundle emission, here we consider the vacuum baths for both the TLS and the optical mode. This is valid because the thermal excitation number is almost zero for typical TLS and bosonic mode at the optical frequency range and room temperature. This treatment is also valid for the superconducting qubit and microwave photons, which have characteristic frequency around $10$ GHz and environment temperature about $15$ mK. However, if the bosonic mode is a mechanical resonator with typical resonance frequency $10$ MHz-$100$ MHz, then a heat bath of the bosonic mode should be considered.
\begin{figure*}[tbp]
	\center
	\includegraphics[width=0.9 \textwidth]{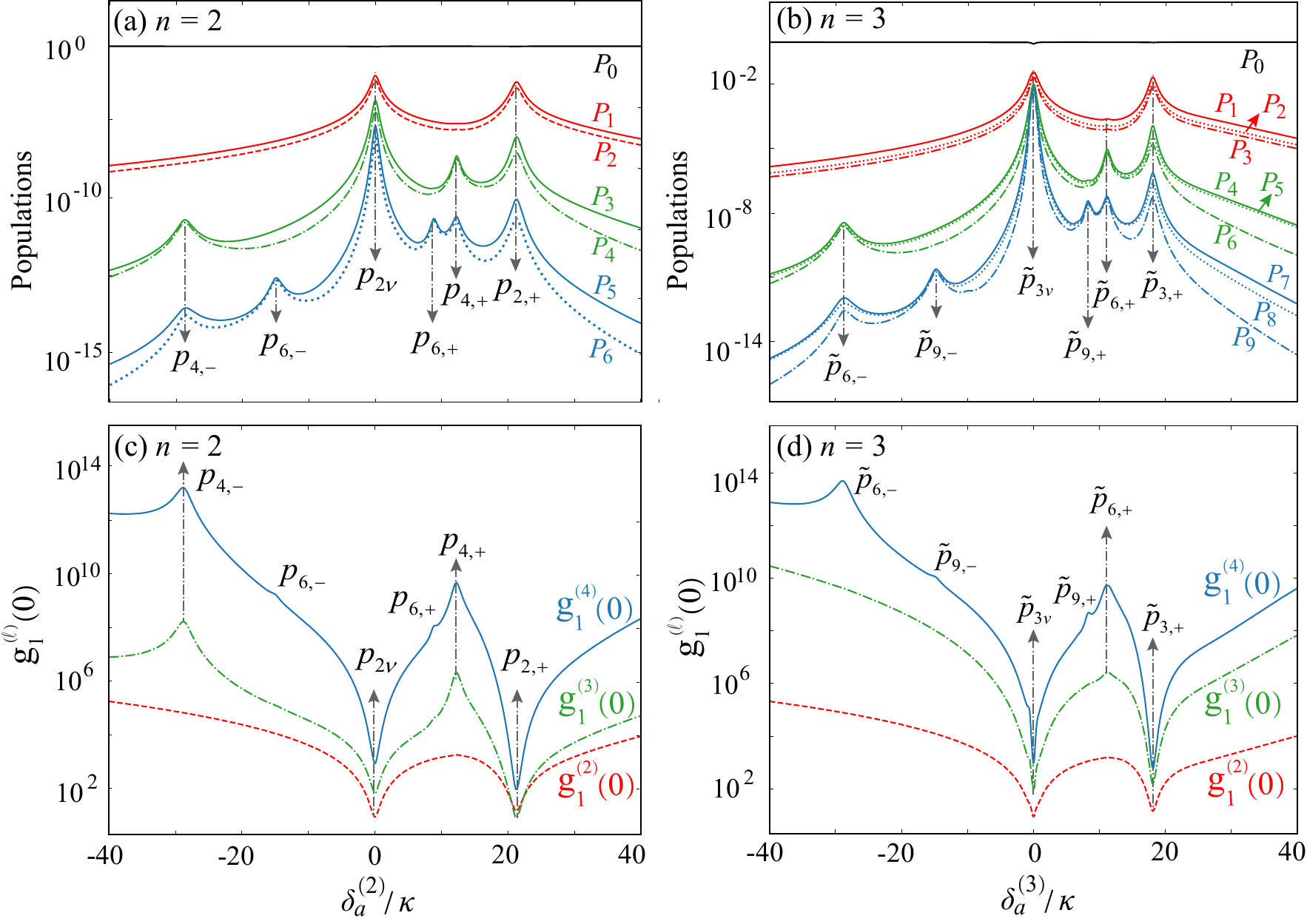}
	\caption{Photon-number distributions $P_{m}$ as functions of the detuning $\delta_{a}^{(n)}/\kappa$ for (a) $m=0$-$6$ and (b) $m=0$-$9$. The standard equal-time $\ell$th-order correlation functions $\mathrm{g}^{(\ell)}_{1}(0)$ as functions of the detuning $\delta_{a}^{(n)}/\kappa$ at (c,~d) $\ell=2,3$, and $4$. Here, the parameters used are (a,~c)  $n=2$, $\Delta^{(2)}/\kappa=-49.5$, and $\Omega_{L}/\kappa=21$, and (b,~d) $n=3$, $\Delta^{(3)}/\kappa=-79.5$, and $\Omega_{L}/\kappa=24$. Other common parameters are $\gamma/\kappa=0.1$, $\gamma_{\phi}=0$, and $J/\kappa=0.3$.}
	\label{Fig3}
\end{figure*}

To study the quantum statistics of the $n$-photon JC model, we analyze the equal-time $\ell$th-order correlation function of the optical mode~\cite{scully_zubairy_1997}, which is defined by
\begin{equation}
\mathrm{g}^{(\ell)}_{1}(0)=\frac{\langle a^{\dagger \ell}a^{\ell} \rangle }{\langle a^{\dagger}a\rangle ^{\ell}}=\frac{\textrm{Tr}(a^{\dagger \ell}a^{\ell}\rho_{\textrm{ss}})}{[\textrm{Tr}(a^{\dagger}a\rho_{\textrm{ss}})]^{\ell}},
\end{equation}
where $\rho_\text{ss}$ is the steady-state density operator of the system [the steady-state solution of Eq.~(\ref{eq:21})]. In addition, based on the steady-state density operator $\rho_\text{ss}$, we can calculate the photon number distributions $P_{m}=\text{Tr}(|m\rangle\langle m|\rho_\text{ss})$. Hence, the equal-time $\ell$th-order correlation function can be expressed as $\mathrm{g}^{(\ell)}_{1}(0)=[\sum_{m=\ell}^{\infty}m!P_{m}/(m-\ell)!]/[(\sum_{m=0}^{\infty}m!P_{m})^{\ell}]$. In the case of few photons, we can qualitatively analyze the effect of the resonance peaks on the equal-time $\ell$th-order correlation function based on the above expression. Concretely, in our following discussions, we will consider the two- and three-photon JC models, i.e., $n=2$ and $3$ in Eq.~(\ref{1}). In Fig.~\ref{Fig3}(a), we show the photon number distributions $P_{m}$~($m=0$-$6$) in the two-photon JC model as functions of the detuning $\delta_{a}^{(2)}/\kappa$. Here, we find that for the two-photon processes, the populations $P_{f}$ and $P_{f-1}$ are close to each other for $f=2,4,~\text{and}~6$, which means the strong correlation of two photons. We also see that the two peaks in $P_{1}$ and $P_{2}$ (red curves) are, respectively, located at $\delta_{a}^{(2)}/\kappa=0$ and $\delta_{a}^{(2)}=-{[(\Delta^{(2)})^{2}+4\Omega_{L}^{2}}]/({4\Delta^{(2)}})\approx21.28\kappa$, under the omission of the frequency shift in $\delta_{a}^{(2)}$ induced by the two-photon JC coupling. By analyzing the eigenenergy levels of the Hamiltonian $H_{0}$ in Eq.~(\ref{H_{0}}), we find that the locations of these peaks in the curves of $P_{1}$ and $P_{2}$ are determined by the $2\nu$-photon~($\nu = 1,~\mu$) resonance transitions $\vert0\rangle\vert +\rangle\leftrightarrow\vert 2\nu\rangle\vert +\rangle$ ($\vert0\rangle\vert -\rangle\leftrightarrow\vert 2\nu\rangle\vert -\rangle$) and the two-photon resonance transition $\vert0\rangle\vert +\rangle\leftrightarrow\vert 2\rangle\vert- \rangle$. To be more clearer, we mark these peaks in the curves of $P_{1}$ and $P_{2}$ as $p_{2\nu}$ and $p_{2,+}$. In the populations $P_{3}$ and $P_{4}$ (green curves), we can observe that there are four peaks located at $\delta_{a}^{(2)}/\kappa=0$, $\delta_{a}^{(2)}/\kappa\approx21.28$, and $ \delta_{a,2}^{(2)}=[\Delta^{(2)}\pm2\sqrt{(\Delta^{(2)})^{2}+3\Omega_{L}^{2}}]/6$, respectively. The locations of the two peaks $p_{4,\pm}$ are determined by the four-photon resonance transitions $\vert0\rangle\vert +\rangle\leftrightarrow\vert 4\rangle\vert -\rangle$ and $\vert0\rangle\vert -\rangle\leftrightarrow\vert 4\rangle\vert +\rangle$, while the other two peaks are induced by the $2\nu$-photon~($\nu = 1,~\mu$) resonance transitions [$\vert0\rangle\vert +\rangle\leftrightarrow\vert 2\nu\rangle\vert +\rangle$ ($\vert0\rangle\vert -\rangle\leftrightarrow\vert 2\nu\rangle\vert -\rangle$)] and the two-photon resonance transition ($\vert0\rangle\vert +\rangle\leftrightarrow\vert 2\rangle\vert- \rangle$). Hence, the locations of these two peaks are the same as those in the curves of $P_{1}$ and $P_{2}$. Similarly, in the populations $P_{5}$ and $P_{6}$ (blue curves), we can observe that there are six peaks located at $\delta_{a}^{(2)}/\kappa=0$, $\delta_{a}^{(2)}/\kappa\approx21.28$, $
\delta_{a,2}^{(2)}=[\Delta^{(2)}\pm2\sqrt{(\Delta^{(2)})^{2}+3\Omega_{L}^{2}}]/6$, and $\delta_{a,3}^{(2)}=[\Delta^{(2)}\pm\sqrt{9(\Delta^{(2)})^{2}+32\Omega_{L}^{2}}]/16$, respectively. Except for those four peaks at the same locations as the four peaks in the curves of $P_{3}$ and $P_{4}$, the locations of the two peaks $p_{6,\pm}$ are determined by the six-photon resonance transitions $\vert0\rangle\vert +\rangle\leftrightarrow\vert 6\rangle\vert -\rangle$ and $\vert0\rangle\vert -\rangle\leftrightarrow\vert 6\rangle\vert +\rangle$.

In Fig.~\ref{Fig3}(b), we show the photon number distributions $P_{m}$ ($m=0$-$9$) in the three-photon JC model as functions of the detuning $\delta_{a}^{(3)}/\kappa$.
For the three-photon process, the populations $P_{f}$, $P_{f-1}$,~and~$P_{f-2}$ are close to each other for $f=3$, 6, and 9, which indicates the strong correlation of three photons. Similarly, the six peaks ($\tilde{p}_{3\nu},~\tilde{p}_{3,+},~\tilde{p}_{6,\pm},~\tilde{p}_{9,\pm}$) correspond to the multiple-photon resonance transitions $\vert0\rangle\vert +\rangle\leftrightarrow\vert 3\nu\rangle\vert +\rangle$ ($\vert0\rangle\vert -\rangle\leftrightarrow\vert 3\nu\rangle\vert -\rangle$)~($\nu = 1,~\mu$), $\vert0\rangle\vert +\rangle\leftrightarrow\vert 3\rangle\vert- \rangle$, $\vert0\rangle\vert +\rangle\leftrightarrow\vert 6\rangle\vert -\rangle$ ($\vert0\rangle\vert -\rangle\leftrightarrow\vert 6\rangle\vert +\rangle$), and $\vert0\rangle\vert +\rangle\leftrightarrow\vert 9\rangle\vert -\rangle$ ($\vert0\rangle\vert -\rangle\leftrightarrow\vert 9\rangle\vert +\rangle$), respectively. The locations of these six peaks can be obtained as $\delta_{a}^{(3)}/\kappa=0$, $\delta_{a}^{(3)}=-{[(\Delta^{(3)})^{2}+4\Omega_{L}^{2}}]/({6\Delta^{(2)}})$, $\delta_{a,2}^{(3)}=[\Delta^{(3)}\pm2\sqrt{(\Delta^{(3)})^{2}+3\Omega_{L}^{2}}]/9$, and $\delta_{a,3}^{(3)}=[\Delta^{(3)}\pm\sqrt{9(\Delta^{(3)})^{2}+32\Omega_{L}^{2}}]/24$.

In Fig.~\ref{Fig3}(c), we show the equal-time $\ell$th-order correlation functions $\mathrm{g}^{(\ell )}_{1}(0)$ ($\ell =2, 3$, and $4$) in the two-photon JC model as functions of the detuning $\delta_{a}^{(2)}/\kappa$. It can be seen that the values of the correlation functions $\mathrm{g}^{(\ell )}_{1}(0)$ are larger than 1 in all the parameter region, which indicates the strong correlation (super-Poisson distribution) of the emitted photons.  In the curve of $\mathrm{g}^{(2)}_{1}(0)$ (red dashed curve), we find two dips [$p_{2\nu}~(\nu=1,~\mu)$ and $p_{2,+}$] rather than two bunching peaks at both $\delta_{a}^{(2)}=0$ and $\delta_{a}^{(2)}/\kappa\approx21.28$. The dip at $\delta_{a}^{(2)}=0$ corresponds to the 2$\nu$-photon resonant transitions $\vert0\rangle\vert-\rangle\leftrightarrow\vert2\nu\rangle\vert-\rangle$ and $\vert0\rangle\vert+\rangle\leftrightarrow\vert2\nu\rangle\vert+\rangle$, and the dip at $\delta_{a}^{(2)}/\kappa\approx21.28$ corresponds to the two-photon resonant transition $\vert0\rangle\vert+\rangle\leftrightarrow\vert2\rangle\vert-\rangle$. This means that the system enters the regime of two-photon bundle emission at the two-photon resonant transition. In the case of $\delta_{a}^{(2)}=0$, we find from Fig.~\ref{Fig1}(c) that the two-photon resonant transitions $\vert2\rangle\vert-\rangle\leftrightarrow\vert4\rangle\vert-\rangle$ and $\vert2\rangle\vert+\rangle\leftrightarrow\vert4\rangle\vert+\rangle$ can be realized, which indicates super-Poisson distribution of two-photon bundle. In the case of $\delta_{a}^{(2)}/\kappa\approx21.28$, we find from Fig.~\ref{Fig1}(c) that the two-photon transitions $\vert2\rangle\vert-\rangle\leftrightarrow\vert4\rangle\vert\pm\rangle$ cannot be realized, which indicates sub-Poisson distribution of two-photon bundle. Except for the two dips at the same locations as those in the curves of $\mathrm{g}^{(2)}_{1}(0)$, we can see that the two peaks ($p_{4,\pm}$) in the curve of $\mathrm{g}^{(3)}_{1}(0)$ (green dash-dotted curve), at the locations $\delta_{a,2}^{(2)}=[\Delta^{(2)}\pm2\sqrt{(\Delta^{(2)})^{2}+3\Omega_{L}^{2}}]/6$, correspond to the four-photon resonant transitions $\vert0\rangle\vert+\rangle\leftrightarrow\vert4\rangle\vert-\rangle$ and $\vert0\rangle\vert-\rangle\leftrightarrow\vert4\rangle\vert+\rangle$. In the curve of $\mathrm{g}^{(4)}_{1}(0)$ (blue solid curves), we see there are four peaks located at $\delta_{a,2}^{(2)}=[\Delta^{(2)}\pm2\sqrt{(\Delta^{(2)})^{2}+3\Omega_{L}^{2}}]/6$, and $\delta_{a,3}^{(2)}=
[\Delta^{(2)}\pm\sqrt{9(\Delta^{(2)})^{2}+32\Omega_{L}^{2}}]/16$, respectively. Except the two peaks and two dips at the same locations as the two peaks and two dips in the curves of $\mathrm{g}^{(3)}_{1}(0)$, the locations of the two peaks $p_{6,\pm}$ are determined by the six-photon resonance transitions $\vert0\rangle\vert +\rangle\leftrightarrow\vert 6\rangle\vert -\rangle$ and $\vert0\rangle\vert -\rangle\leftrightarrow\vert 6\rangle\vert +\rangle$.

In Fig.~\ref{Fig3}(d), we plot the correlation functions $\mathrm{g}^{(\ell )}_{1}(0)$ ($\ell =2, 3$, and $4$) in the three-photon JC model as functions of the detuning $\delta_{a}^{(3)}/\kappa$. Similarly, the two dips [$\tilde{p} _{3\nu}~(\nu=1,~\mu)$ and $\tilde{p}_{3,+}$] for each $\mathrm{g}^{(\ell )}_{1}(0)$ can be observed at the 3$\nu$-photon resonant transitions  $\vert0\rangle\vert-\rangle\leftrightarrow\vert3\nu\rangle\vert-\rangle$ ($\vert0\rangle\vert+\rangle\leftrightarrow\vert3\nu\rangle\vert+\rangle$) and the three-photon resonant transitions  $\vert0\rangle\vert+\rangle\leftrightarrow\vert3\rangle\vert-\rangle$, i.e., $\delta_{a}^{(3)}=0$ and $\delta_{a}^{(3)}=-{[(\Delta^{(3)})^{2}+4\Omega_{L}^{2}}]/({6\Delta^{(3)}})\approx18.08\kappa$. This means that the system enters the regime of three-photon bundle emission at the three-photon resonant transition. In addition, we observe four peaks $(\tilde{p}_{6,\pm},~\tilde{p}_{9,\pm})$ in the correlation functions $\mathrm{g}^{(4 )}_{1}(0)$ (blue solid curve), which correspond to the multiple-photon resonant transitions $\vert0\rangle\vert+\rangle\leftrightarrow\vert6\rangle\vert-\rangle$, $\vert0\rangle\vert-\rangle\leftrightarrow\vert6\rangle\vert+\rangle$, $\vert0\rangle\vert+\rangle\leftrightarrow\vert9\rangle\vert-\rangle$, and $\vert0\rangle\vert-\rangle\leftrightarrow\vert9\rangle\vert+\rangle$.
\begin{figure}
	\center
	\includegraphics[width=0.47 \textwidth]{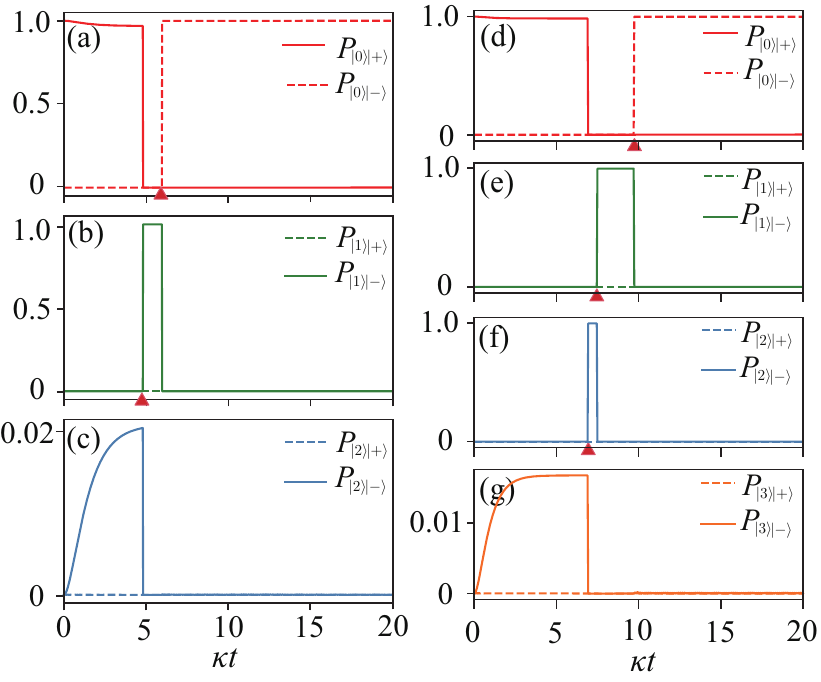}
	\caption{A part of the quantum trajectory of the state populations $P_{\vert m\rangle\vert\pm\rangle}$ ($m=0,1,2,~\text{and}~3$) at (a-c) $n=2$, $\Delta^{(2)}/\kappa=-49.5$, $\Omega_{L}/\kappa=21$, and $\delta_{a}^{(2)}=-[(\Delta^{(2)})^{2}+4\Omega_{L}^{2}]/({4\Delta^{(2)}})$, showing the two-photon bundle emission; (d-g) $n=3$, $\Delta^{(3)}/\kappa=-79.5$, $\Omega_{L}/\kappa=24$, and $\delta_{a}^{(3)}=-[(\Delta^{(3)})^{2}+4\Omega_{L}^{2}]/({6\Delta^{(3)}})$, showing the three-photon bundle emission. Other parameters used are $\gamma/\kappa=0.1$, $\gamma_{\phi}=0$, and $J/\kappa=0.3$.}
	\label{Fig4}
\end{figure}

To exhibit the multiple-photon bundle emission process more clearly, we employ a quantum Monte Carlo simulation to track the individual quantum trajectories of the system. Figures~\ref{Fig4}(a-c) show a short duration of a quantum trajectory of the state populations $P_{\vert m\rangle\vert\pm\rangle}(t)$ $(m=0, 1,$ and $2)$ when $n=2$ and $\delta_{a}^{(2)}/\kappa\approx21.28$. Here we consider that the system is initially in the state $\vert0\rangle\vert+\rangle$. Under the condition of two-photon resonant driving, the two-photon state $\vert2\rangle\vert-\rangle$ is occupied with a probability $0.02$ at time $\kappa t\approx4.8$, while the one-photon state populations $P_{\vert1\rangle\vert\pm\rangle}$ is zero. The dissipation of the optical mode causes the emission of the first photon [indicated by the red triangle in Fig.~\ref{Fig4}(b)], and the wave function of the system collapses to the one-photon state $\vert1\rangle\vert-\rangle$ with almost unit probability. Immediately, the second photon is emitted within the cavity lifetime, and the wave function of the system collapses to the zero-photon state $\vert0\rangle\vert-\rangle$ [indicated by the red triangle in Fig.~\ref{Fig4}(a)], resulting in the two-photon bundle emission (the emission of two strongly-correlated photons). It is worth noting that the relation $c_{-}\gg c_{+}$ is satisfied in the cases of $\Omega_{L}/\kappa=21$, $\Delta^{(2)}/\kappa=-49.5$, and $\delta_{a}^{(2)}/\kappa\approx21.28$, then the eigenstates $\vert-\rangle$ and $\vert+\rangle$ of the TLS can be approximately reduced to the bare states $\vert e\rangle$ and $\vert g\rangle$, respectively. This means that the system is in the state $\vert0\rangle\vert e\rangle$ after the two-photon bundle emission. Due to the dissipation of the TLS, the system goes back to the state $\vert0\rangle\vert g\rangle\approx\vert0\rangle\vert +\rangle$, as the starting state of the next emission of two strongly-correlated photons. In Figs.~\ref{Fig4}(d-g), we show a short duration of a quantum trajectory of the state populations $P_{\vert m\rangle\vert\pm\rangle}(t)$ $(m=0, 1, 2,~\text{and}~3)$ when $n=3$ and $\delta_{a}^{(3)}/\kappa\approx18.08$. It can be seen that the three-photon bundle emission can be realized in the presence of the dissipation of the system.
\begin{figure}
\center
\includegraphics[width=0.47 \textwidth]{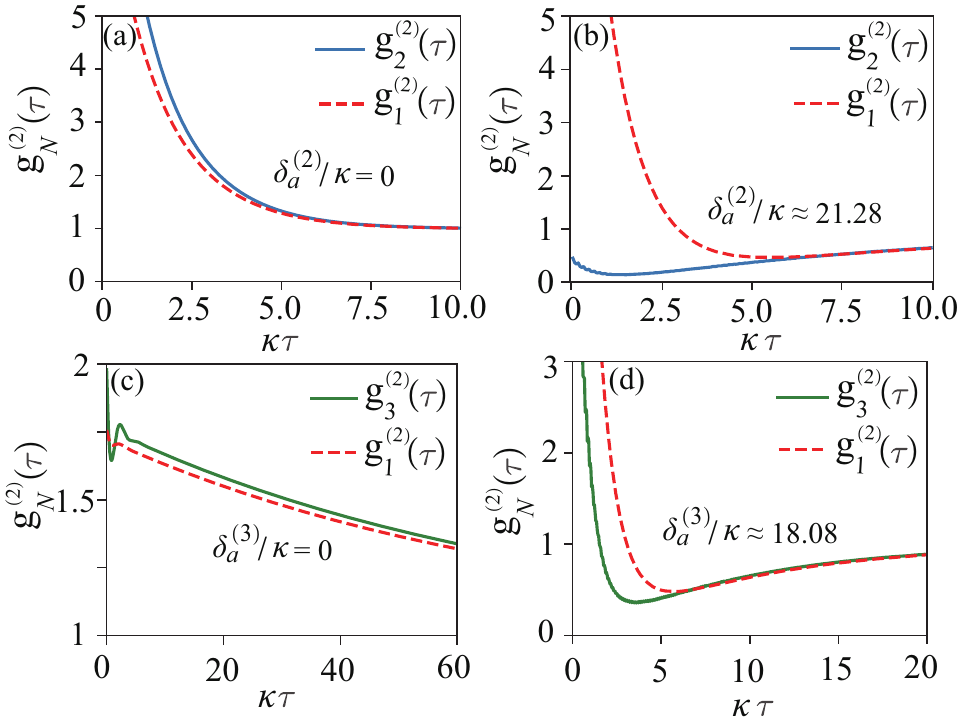}
\caption{The time-delay second-order correlation functions $\mathrm{g}_{N}^{(2)}(\tau)$ as functions of the scaled evolution time $\kappa\tau$ for $n=2$ at (a) $\delta_{a}^{(2)}=0$ and (b) $\delta_{a}^{(2)}=-[(\Delta^{(2)})^{2}+4\Omega_{L}^{2}]/(4\Delta^{(2)})$, and for $n=3$ at (c) $\delta_{a}^{(3)}=0$ and (d)  $\delta_{a}^{(3)}=-[(\Delta^{(3)})^{2}+4\Omega_{L}^{2}]/(6\Delta^{(3)})$. The red dashed curves, blue solid curves, and green solid curves correspond to $\mathrm{g}_{1}^{(2)}(\tau)$, $\mathrm{g}_{2}^{(2)}(\tau)$, and $\mathrm{g}_{3}^{(2)}(\tau)$, respectively. Here, the parameters used are (a,~b) $\Omega_{L}/\kappa=21$ and (c,~d) $\Omega_{L}/\kappa=24$. Other common parameters are $\gamma/\kappa=0.1$, $\gamma_{\phi}=0$, $J/\kappa=0.3$, $\Delta^{(2)}/\kappa=-49.5$, and $\Delta^{(3)}/\kappa=-79.5$.}
\label{Fig5}
\end{figure}

To further characterize the quantum statistical properties between the multiple-photon bundles, we calculate the generalized $\ell$th-order correlation function of $N$-photon bundle~\cite{10.1038/nphoton.2014.114}
\begin{equation}
\mathrm{g}_{N}^{(\ell)}(t_{1},\cdots,t_{\ell})=\frac{\langle \mathcal{T}_{-}\{ \Pi_{i=1}^{\ell}a^{\dagger N}(t_{i})\} \mathcal{T}_{+}\{ \Pi_{i=1}^{\ell}a^{N}(t_{i})\} \rangle }{\Pi_{i=1}^{\ell}\langle a^{\dagger N}a^{N}\rangle (t_{i})},\label{gn}
\end{equation}
where $\mathcal{T}_{\pm}$ denotes the time-ordering operator. The operators $a^{\dagger N}$ ($a^{N}$) in Eq.~({\ref{gn}}) represent the simultaneous creation (annihilation) of $N$ photons, which corresponds to the process of $N$-photon bundle emission. Concretely, we analyze the time-delay second-order correlation function of $N$-photon bundle
\begin{equation}
\mathrm{g}_{N}^{(2)}(\tau)=\frac{\langle a^{\dagger N}(0)a^{\dagger N}(\tau)a^{N}(\tau)a^{N}(0)\rangle}{\langle (a^{\dagger N}a^{N})(0)\rangle\langle (a^{\dagger N}a^{N})(\tau)\rangle},\label{gn2}
\end{equation}
where $\tau$ is delayed time. In the case of $N=1$, the generalized time-delay second-order correlation function $\mathrm{g}_{N}^{(2)}(\tau)$ goes back to the standard time-delay second-order correlation function $\mathrm{g}_{1}^{(2)}(\tau)=\langle a^{\dagger}(0)a^{\dagger}(\tau)a(\tau)a(0)\rangle/[\langle a^{\dagger}a(0)\rangle\langle a^{\dagger}a(\tau)\rangle]$.

To study the statistical properties of two-photon bundle in the two-photon JC model, we show the correlation functions $\mathrm{g}_{1}^{(2)}(\tau)$ and $\mathrm{g}_{2}^{(2)}(\tau)$ as functions of the delayed time $\kappa\tau$ at $\delta_{a}^{(2)}=0$ in Fig.~\ref{Fig5}(a) and $\delta_{a}^{(2)}/\kappa\approx21.28$ in Fig.~\ref{Fig5}(b). Here the delayed time $\tau$ is larger than $\tau^{[N]}_{\text{min}}=\sum_{m=1}^{N}{(1/m\kappa)}$, and $\mathrm{g}_{N}^{(2)}(\tau^{[N]}_{\text{min}})$ can be approximately regarded as zero-time-delay correlation function for the case of multiple-photon bundle. The reason is that the generalized second-order correlation function $\mathrm{g}_{N}^{(2)}(\tau)$ is ill-defined in a short temporal window of width $\tau^{[N]}_{\text{min}}$~\cite{10.1038/nphoton.2014.114,bin2020paritysymmetryprotected}. It can be seen from Fig.~\ref{Fig5}(a) that $\mathrm{g}_{1}^{(2)}(0)>\mathrm{g}_{1}^{(2)}(\tau)$ and $\mathrm{g}_{2}^{(2)}(\tau^{[2]}_{\text{min}})>\mathrm{g}_{2}^{(2)}(\tau)$, which corresponds to the bunching of photon pairs at $\delta_{a}^{(2)}=0$. In Fig.~\ref{Fig5}(b), we observe that the correlation functions satisfy $\mathrm{g}_{1}^{(2)}(0)>\mathrm{g}_{1}^{(2)}(\tau)$ and $\mathrm{g}_{2}^{(2)}(\tau^{[2]}_{\text{min}})<\mathrm{g}_{2}^{(2)}(\tau)$. This indicates that the antibunching of photon pairs can be realized in the case of $\delta_{a}^{(2)}/\kappa\approx21.28$. In Figs.~\ref{Fig5}(c) and~\ref{Fig5}(d), the correlation functions $\mathrm{g}_{1}^{(2)}(\tau)$ and $\mathrm{g}_{3}^{(2)}(\tau)$ in the three-photon JC model are plotted as functions of the delayed time $\kappa\tau$. At $\delta_{a}^{(3)}=0$, it can be seen from Fig.~\ref{Fig5}(c) that $\mathrm{g}_{1}^{(2)}(0)>g_{1}^{(2)}(\tau)$ and $\mathrm{g}_{3}^{(2)}(\tau^{[3]}_{\text{min}})>\mathrm{g}_{3}^{(2)}(\tau)$, corresponding to the bunching of three strongly-correlated photons. At $\delta_{a}^{(3)}/\kappa\approx18.08$, we find from Fig.~\ref{Fig5}(d) that $\mathrm{g}_{1}^{(2)}(0)>\mathrm{g}_{1}^{(2)}(\tau)$ and $\mathrm{g}_{3}^{(2)}(\tau^{[3]}_{\text{min}})<\mathrm{g}_{3}^{(2)}(\tau)$, which indicate the antibunching of three strongly-correlated photons. Hence, the statistical properties of multiple-photon bundle can be tuned by choosing the appropriate resonant transitions, and then the $n$-photon JC system can behave as a multiple-photon gun.

\begin{figure}
\center
\includegraphics[width=0.47 \textwidth]{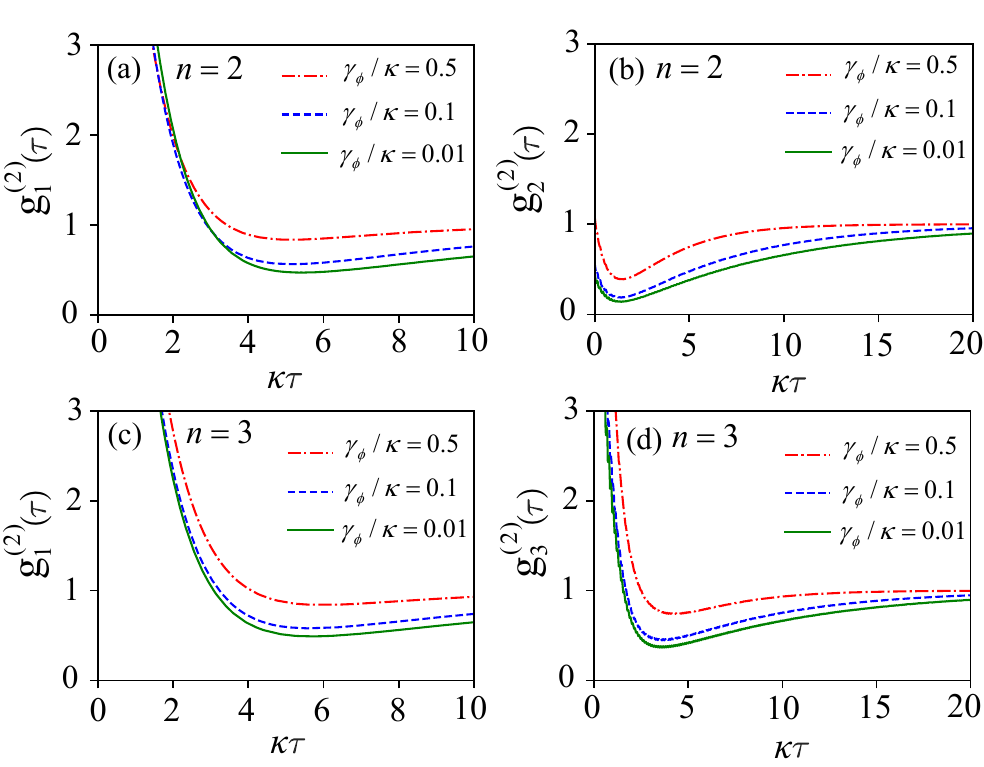}
\caption{(a) $\mathrm{g}_{1}^{(2)}(\tau)$ and (b) $\mathrm{g}_{2}^{(2)}(\tau)$ as functions of the scaled evolution time $\kappa\tau$ at various values $\gamma_{\phi}/\kappa=(0.01, 0.1, 0.5)$ for $n=2$. (c) $\mathrm{g}_{1}^{(2)}(\tau)$ and (d) $\mathrm{g}_{3}^{(2)}(\tau)$ as functions of the scaled evolution time $\kappa\tau$ at various values $\gamma_{\phi}/\kappa=(0.01, 0.1, 0.5)$ for $n=3$. Here, the parameters used are (a,~b) $\delta_{a}^{(2)}=-[(\Delta^{(2)})^{2}+4\Omega_{L}^{2}]/(4\Delta^{(2)})\approx21.28\kappa$ and $\Omega_{L}/\kappa=21$, and (c,~d) $\delta_{a}^{(3)}=-[(\Delta^{(3)})^{2}+4\Omega_{L}^{2}]/(6\Delta^{(3)})\approx18.08\kappa$ and $\Omega_{L}/\kappa=24$. Other common parameters are $\gamma/\kappa=0.1$, $J/\kappa=0.3$, $\Delta^{(2)}/\kappa=-49.5$, and $\Delta^{(3)}/\kappa=-79.5$.}
\label{Fig6}
\end{figure}

In the above discussions, we neglected the effect of the pure dephasing of the TLS. Below, we will discuss the influence of the pure dephasing on the time-delayed second-order correlation functions. In Figs.~\ref{Fig6}(a) and~\ref{Fig6}(b), we show the correlation functions $\mathrm{g}_{1}^{(2)}(\tau)$ and $\mathrm{g}_{2}^{(2)}(\tau)$ as functions of the scaled evolution time $\kappa\tau$ at various values of the pure dephasing rate $\gamma_{\phi}/\kappa=(0.01, 0.1, 0.5)$ for $n=2$ and $\delta_{a}^{(2)}/\kappa\approx21.28$. It can be seen that the values of $\mathrm{g}_{1}^{(2)}(\tau)$ and $\mathrm{g}_{2}^{(2)}(\tau)$ raise as the pure dephasing rate increases. In particular, we observe that $\mathrm{g}_{1}^{(2)}(0)>\mathrm{g}_{1}^{(2)}(\tau)$ and $\mathrm{g}_{2}^{(2)}(\tau^{[2]}_{\text{min}})>\mathrm{g}_{2}^{(2)}(\tau)$ in the presence of the pure dephasing, which means the antibunching of photon pairs can be realized in the presence of the pure dephasing of the TLS. In Figs.~\ref{Fig6}(c) and~\ref{Fig6}(d), the correlation functions $\mathrm{g}_{1}^{(2)}(\tau)$ and $\mathrm{g}_{3}^{(2)}(\tau)$ are plotted as functions of $\kappa\tau$ when $\gamma_{\phi}/\kappa=(0.01, 0.1, 0.5)$ for $n=3$ and $\delta_{a}^{(3)}/\kappa\approx18.08$. Figures~\ref{Fig6}(c) and~\ref{Fig6}(d) show that $\mathrm{g}_{1}^{(2)}(0)>\mathrm{g}_{1}^{(2)}(\tau)$ and $\mathrm{g}_{3}^{(2)}(\tau^{[3]}_{\text{min}})<\mathrm{g}_{3}^{(2)}(\tau)$. This indicates that the antibunching of three strongly-correlated photons can be realized even when the pure dephasing rate $\gamma_{\phi}$ of the TLS is larger than the decay rate $\gamma$.

\section{Discussions}\label{sectionIV}

In the above sections, we consider the multiple-photon bundle emission in the Mollow regime, in which the driving magnitude is much stronger than the $n$-photon JC coupling strength. An interesting question arised here is whether we can implement the multiple-photon bundle emission in the $n$-photon JC regime, in which the $n$-photon JC couping strength $J$ is much stronger than the driving amplitude $\Omega_{L}$ and then the driving term can be treated as a perturbation. Below we present some discussions on the multiple-photon bundle emission in the $n$-photon JC regime. For the rotated $n$-photon JC Hamiltonian
	\begin{equation}\label{H0'}
		\tilde{H}_{0}=\delta_{a}^{(n)}a^{\dagger}a+\delta_{\sigma}\sigma_{+}\sigma_{-}+J(a^{\dagger n}\sigma_{-}+\sigma_{+}a^{n}),
	\end{equation}
its eigensystem reads $\tilde{H}_{0}\vert g,m\rangle=m\delta_{a}^{(n)}\vert g,m\rangle\,(0\le m<n)$ and $\tilde{H}_{0}\vert \varepsilon_{m,\pm}\rangle=E_{m,\pm}\vert \varepsilon_{m,\pm}\rangle\,(m\ge n)$, where the eigenvalues and eigenstates are given by $E_{m,\pm}=(m-n/2)\delta_{a}^{(n)}+[\delta_{\sigma}\pm\Omega_{m}(\Delta^{(n)})]/2$ and $\vert\varepsilon_{m,\pm}\rangle=C_{\mp}^{[m]}\vert g,m\rangle\pm C_{\pm}^{[m]}\vert e,m-n\rangle$, with the introduced variables $C_{\pm}^{[m]}=\sqrt{[\Omega_{m}(\Delta^{(n)})\pm\Delta^{(n)}]/2\Omega_{m}(\Delta^{(n)})}$ and $\Omega_{m}(\Delta^{(n)})=\sqrt{(\Delta^{(n)})^{2}+4J^{2}m!/(m-n)!}$. By applying the condition~$|\Delta^{(n)}| \gg J\sqrt{m!/(m-n)!}$, we can obtain the relation
	$C_{-}^{[m]}\gg C_{+}^{[m]}~(C_{+}^{[m]}\gg C_{-}^{[m]})$, and thus the eigenstates $|\varepsilon_{m,+}\rangle$ and $|\varepsilon_{m,-}\rangle$ can be approximately reduced to the bare states $|g,m\rangle~(|e,m-n\rangle)$ and $-|e,m-n\rangle~(|g,m\rangle)$, respectively.

We consider the initial state  $|g,0\rangle$ of the system and choose the resonant transition $|g,0\rangle\leftrightarrow|\varepsilon_{2n,-}\rangle~(-\vert e,n\rangle)$, then the system can be approximately restricted into the subspace with the two basis states $\{\vert g,0\rangle,\vert
 e,n\rangle\}$, and the super-Rabi oscillation can occur between these two states. We note that the effective resonance frequency of the super-Rabi oscillation between the zero-photon state $|g,0\rangle$ and the $n$-photon state $|e,n\rangle$ can be analytically obtained as
\begin{equation}
	\tilde{\Omega}_{\textrm{eff}}^{(n)}=\frac{J\sqrt{n!}\Omega_{L}^{2}}{n\delta_{a}^{(n)}\delta
		_{\sigma}-n!J^2}.
\end{equation}	
We checked that the analytical results match with the numerical results well under proper resonant conditions. We should point out that the resonant condition for the super-Rabi oscillation in the $n$-photon JC regime is extreme. It is found that the precision of the detuning $\delta_{a}^{(n)}$ for $n=2~\text{and}~3$ should be chosen up to $10^{-6}$ such that a good super-Rabi oscillation can be kept. For example, we need to take $\delta_{a}^{(2)}/J=41.268234~\textrm{and}~\delta_{a}^{(3)}/J=44.242035$ under the parameters $\Omega_{L}/J=0.1$, $\Delta^{(2)}/J=-165$, and $\Delta^{(3)}/J=-265$. In the Mollow regime, however, a good super-Rabi oscillation can be obtained when the precision of $\delta_{a}^{(n)}$ is taken up to $10^{-1}$.

We also analyzed the bare state populations and the standard higher-order correlation functions in this regime. However, the result shows that the $n$-photon state population $P_{|e,n\rangle }$ is extremely small and the correlation function shows a sharp peak under $2n$-excitation resonance conditions for $n=2~\text{and}~3$.
This indicates that it is difficult to generate $n$-photon bundle emission in $n$-photon JC regime under the used parameters.

Finally, we present some discussions on the experimental implementation of the current scheme. In this scheme, there are two important factors: the $n$-photon JC interaction and the atomic driving. In our discussions, we used the terms of photons and TLS, but did not prescribe a specific experimental setup. In this sense, the physical mechanism proposed in this work is general, and hence it can be implemented with physical systems in which both the $n$-photon JC interaction and the driving on the TLS can be realized. Currently, many theoretical proposals have been proposed for implementation of the $n$-excitation JC interaction with either trapped-ion systems~\cite{PhysRevA.52.4214} or superconducting quantum circuits~\cite{GU20171,Menard2022Apr,chen2017Singlephotondriven}. In the trapped-ion systems~\cite{PhysRevA.52.4214}, the excitations are phonons, and hence the $n$-phonon JC interaction can be realized by considering the red-sideband transition and the $n$-phonon resonance case between the ion internal electronic state and its vibration motion. For the $n$-photon JC model, we note that the $n$-photon processes have been realized in both coupled atom-light systems~\cite{PhysRevLett.59.1899,PhysRevLett.68.464} and superconducting quantum circuits~\cite{chen2017Singlephotondriven}. In addition, we want to mention that the $n$-photon parametric process has been recently realized in a superconducting circuit~\cite{Menard2022Apr}.

Concerning the physical parameters, we also present some analyses to show that the present work should be reliable by current experimental conditions and that the starting $n$-photon JC model is valid. Concretly, in this scheme, the considered $n$-photon JC coupling works in the weak-coupling regime, and the TLS is strongly driven to enter the Mollow regime. In our simultations, we set $J/\kappa=0.3$ and $\Omega_{L}/\kappa=21$ or 24. These values match the weak coupling regime and strong driving condition and they are experimentally accessible based on the current experimental techniques in cavity-QED and circuit-QED setups~\cite{Haroche2006Oct,RevModPhys.73.565,RevModPhys.93.025005}. In particular, the decay rate of the cavity field should be much larger than that of the TLS, for example $\gamma/\kappa=0.1$, such that the bundle emission can occur. This requires that the system works in the bad-cavity regime, which is also a typical parameter regime in cavity-QED. In typical experimental systems in quantum optics, the weak interaction between the photons and the TLS, the strong driving of the TLS, and the bad cavity decay can be implemented with current experimental conditions. In addition, the resonance condition of the multiple photon process can be achieved by tuning the driving frequency and amplitude. Therefore, this scheme should be within the reach of current or near-future experimental techniques in cavity-QED and circuit-QED platforms~\cite{Haroche2006Oct,RevModPhys.73.565,RevModPhys.93.025005}.

\section{Conclusion}\label{sectionV}

In conclusion, we have studied the multiple-photon  bundle emission in the $n$-photon JC model. We have considered the Mollow regime such that a clear super-Rabi oscillation between the zero-photon state and the $n$-photon state can occur. The multiple-photon  bundle emission has been confirmed by investigating the photon number populations, several correlation functions, and the quantum Monter Carlo simulations. Our work is general and hence this scheme can be used to realize multiple-photon  or -phonon sources. This work will pave the way towards the study of multiple-photon  or -phonon devices and quantum information processing.

\begin{acknowledgments}
J.-Q.L. is supported in part by National Natural Science Foundation of China (Grants No. 12175061, No. 12247105, and No. 11935006) and the Science and Technology Innovation Program of Hunan Province (Grants No. 2021RC4029 and No. 2020RC4047). X.-W.X. is supported by the National Science Foundation of China (Grant No. 12064010), Natural Science Foundation of Hunan Province of China (Grant No. 2021JJ20036), and the Science and Technology Innovation Program of Hunan Province (Grant No. 2022RC1203). J.-F.H. is supported in part by the National Natural Science Foundation of China (Grant No. 12075083) and Natural Science Foundation of Hunan Province of China (Grant No. 2020JJ5345). F.Z. is supported in part by the National Natural Science Foundation of China (Grant No. 12147109) and the China Postdoctoral Science Foundation (Grant No. 2021M700360).
\end{acknowledgments}


\begin{thebibliography}{68}%
\makeatletter
\providecommand \@ifxundefined [1]{%
 \@ifx{#1\undefined}
}%
\providecommand \@ifnum [1]{%
 \ifnum #1\expandafter \@firstoftwo
 \else \expandafter \@secondoftwo
 \fi
}%
\providecommand \@ifx [1]{%
 \ifx #1\expandafter \@firstoftwo
 \else \expandafter \@secondoftwo
 \fi
}%
\providecommand \natexlab [1]{#1}%
\providecommand \enquote  [1]{``#1''}%
\providecommand \bibnamefont  [1]{#1}%
\providecommand \bibfnamefont [1]{#1}%
\providecommand \citenamefont [1]{#1}%
\providecommand \href@noop [0]{\@secondoftwo}%
\providecommand \href [0]{\begingroup \@sanitize@url \@href}%
\providecommand \@href[1]{\@@startlink{#1}\@@href}%
\providecommand \@@href[1]{\endgroup#1\@@endlink}%
\providecommand \@sanitize@url [0]{\catcode `\\12\catcode `\$12\catcode
  `\&12\catcode `\#12\catcode `\^12\catcode `\_12\catcode `\%12\relax}%
\providecommand \@@startlink[1]{}%
\providecommand \@@endlink[0]{}%
\providecommand \url  [0]{\begingroup\@sanitize@url \@url }%
\providecommand \@url [1]{\endgroup\@href {#1}{\urlprefix }}%
\providecommand \urlprefix  [0]{URL }%
\providecommand \Eprint [0]{\href }%
\providecommand \doibase [0]{http://dx.doi.org/}%
\providecommand \selectlanguage [0]{\@gobble}%
\providecommand \bibinfo  [0]{\@secondoftwo}%
\providecommand \bibfield  [0]{\@secondoftwo}%
\providecommand \translation [1]{[#1]}%
\providecommand \BibitemOpen [0]{}%
\providecommand \bibitemStop [0]{}%
\providecommand \bibitemNoStop [0]{.\EOS\space}%
\providecommand \EOS [0]{\spacefactor3000\relax}%
\providecommand \BibitemShut  [1]{\csname bibitem#1\endcsname}%
\let\auto@bib@innerbib\@empty
\bibitem [{\citenamefont {Mu{\~{n}}oz}\ \emph {et~al.}(2014)\citenamefont
  {Mu{\~{n}}oz}, \citenamefont {del Valle}, \citenamefont {Tudela},
  \citenamefont {M{\"u}ller}, \citenamefont {Lichtmannecker}, \citenamefont
  {Kaniber}, \citenamefont {Tejedor}, \citenamefont {Finley},\ and\
  \citenamefont {Laussy}}]{10.1038/nphoton.2014.114}%
  \BibitemOpen
  \bibfield  {author} {\bibinfo {author} {\bibfnamefont {C.~S.}\ \bibnamefont
  {Mu{\~{n}}oz}}, \bibinfo {author} {\bibfnamefont {E.}~\bibnamefont {del
  Valle}}, \bibinfo {author} {\bibfnamefont {A.~G.}\ \bibnamefont {Tudela}},
  \bibinfo {author} {\bibfnamefont {K.}~\bibnamefont {M{\"u}ller}}, \bibinfo
  {author} {\bibfnamefont {S.}~\bibnamefont {Lichtmannecker}}, \bibinfo
  {author} {\bibfnamefont {M.}~\bibnamefont {Kaniber}}, \bibinfo {author}
  {\bibfnamefont {C.}~\bibnamefont {Tejedor}}, \bibinfo {author} {\bibfnamefont
  {J.~J.}\ \bibnamefont {Finley}}, \ and\ \bibinfo {author} {\bibfnamefont
  {F.~P.}\ \bibnamefont {Laussy}},\ }\href {\doibase 10.1038/nphoton.2014.114}
  {\bibfield  {journal} {\bibinfo  {journal} {Nat. Photonics}\ }\textbf
  {\bibinfo {volume} {8}},\ \bibinfo {pages} {550} (\bibinfo {year}
  {2014})}\BibitemShut {NoStop}%
\bibitem [{\citenamefont {Kimble}(2008)}]{Kimble2008Jun}%
  \BibitemOpen
  \bibfield  {author} {\bibinfo {author} {\bibfnamefont {H.~J.}\ \bibnamefont
  {Kimble}},\ }\href {\doibase 10.1038/nature07127} {\bibfield  {journal}
  {\bibinfo  {journal} {Nature}\ }\textbf {\bibinfo {volume} {453}},\ \bibinfo
  {pages} {1023} (\bibinfo {year} {2008})}\BibitemShut {NoStop}%
\bibitem [{\citenamefont {Boto}\ \emph {et~al.}(2000)\citenamefont {Boto},
  \citenamefont {Kok}, \citenamefont {Abrams}, \citenamefont {Braunstein},
  \citenamefont {Williams},\ and\ \citenamefont {Dowling}}]{PhysRevLett2000}%
  \BibitemOpen
  \bibfield  {author} {\bibinfo {author} {\bibfnamefont {A.~N.}\ \bibnamefont
  {Boto}}, \bibinfo {author} {\bibfnamefont {P.}~\bibnamefont {Kok}}, \bibinfo
  {author} {\bibfnamefont {D.~S.}\ \bibnamefont {Abrams}}, \bibinfo {author}
  {\bibfnamefont {S.~L.}\ \bibnamefont {Braunstein}}, \bibinfo {author}
  {\bibfnamefont {C.~P.}\ \bibnamefont {Williams}}, \ and\ \bibinfo {author}
  {\bibfnamefont {J.~P.}\ \bibnamefont {Dowling}},\ }\href {\doibase
  10.1103/PhysRevLett.85.2733} {\bibfield  {journal} {\bibinfo  {journal}
  {Phys. Rev. Lett.}\ }\textbf {\bibinfo {volume} {85}},\ \bibinfo {pages}
  {2733} (\bibinfo {year} {2000})}\BibitemShut {NoStop}%
\bibitem [{\citenamefont {D'Angelo}\ \emph {et~al.}(2001)\citenamefont
  {D'Angelo}, \citenamefont {Chekhova},\ and\ \citenamefont
  {Shih}}]{Two-photon2001}%
  \BibitemOpen
  \bibfield  {author} {\bibinfo {author} {\bibfnamefont {M.}~\bibnamefont
  {D'Angelo}}, \bibinfo {author} {\bibfnamefont {M.~V.}\ \bibnamefont
  {Chekhova}}, \ and\ \bibinfo {author} {\bibfnamefont {Y.}~\bibnamefont
  {Shih}},\ }\href {\doibase 10.1103/PhysRevLett.87.013602} {\bibfield
  {journal} {\bibinfo  {journal} {Phys. Rev. Lett.}\ }\textbf {\bibinfo
  {volume} {87}},\ \bibinfo {pages} {013602} (\bibinfo {year}
  {2001})}\BibitemShut {NoStop}%
\bibitem [{\citenamefont {Gisin}\ \emph {et~al.}(2002)\citenamefont {Gisin},
  \citenamefont {Ribordy}, \citenamefont {Tittel},\ and\ \citenamefont
  {Zbinden}}]{Gisin2002Mar}%
  \BibitemOpen
  \bibfield  {author} {\bibinfo {author} {\bibfnamefont {N.}~\bibnamefont
  {Gisin}}, \bibinfo {author} {\bibfnamefont {G.}~\bibnamefont {Ribordy}},
  \bibinfo {author} {\bibfnamefont {W.}~\bibnamefont {Tittel}}, \ and\ \bibinfo
  {author} {\bibfnamefont {H.}~\bibnamefont {Zbinden}},\ }\href {\doibase
  10.1103/RevModPhys.74.145} {\bibfield  {journal} {\bibinfo  {journal} {Rev.
  Mod. Phys.}\ }\textbf {\bibinfo {volume} {74}},\ \bibinfo {pages} {145}
  (\bibinfo {year} {2002})}\BibitemShut {NoStop}%
\bibitem [{\citenamefont {Giovannetti}\ \emph {et~al.}(2004)\citenamefont
  {Giovannetti}, \citenamefont {Lloyd},\ and\ \citenamefont
  {Maccone}}]{Giovannetti1330}%
  \BibitemOpen
  \bibfield  {author} {\bibinfo {author} {\bibfnamefont {V.}~\bibnamefont
  {Giovannetti}}, \bibinfo {author} {\bibfnamefont {S.}~\bibnamefont {Lloyd}},
  \ and\ \bibinfo {author} {\bibfnamefont {L.}~\bibnamefont {Maccone}},\ }\href
  {\doibase 10.1126/science.1104149} {\bibfield  {journal} {\bibinfo  {journal}
  {Science}\ }\textbf {\bibinfo {volume} {306}},\ \bibinfo {pages} {1330}
  (\bibinfo {year} {2004})}\BibitemShut {NoStop}%
\bibitem [{\citenamefont {Giovannetti}\ \emph {et~al.}(2006)\citenamefont
  {Giovannetti}, \citenamefont {Lloyd},\ and\ \citenamefont
  {Maccone}}]{PhysRevLett.96.010401}%
  \BibitemOpen
  \bibfield  {author} {\bibinfo {author} {\bibfnamefont {V.}~\bibnamefont
  {Giovannetti}}, \bibinfo {author} {\bibfnamefont {S.}~\bibnamefont {Lloyd}},
  \ and\ \bibinfo {author} {\bibfnamefont {L.}~\bibnamefont {Maccone}},\ }\href
  {\doibase 10.1103/PhysRevLett.96.010401} {\bibfield  {journal} {\bibinfo
  {journal} {Phys. Rev. Lett.}\ }\textbf {\bibinfo {volume} {96}},\ \bibinfo
  {pages} {010401} (\bibinfo {year} {2006})}\BibitemShut {NoStop}%
\bibitem [{\citenamefont {Douglas}\ \emph {et~al.}(2016)\citenamefont
  {Douglas}, \citenamefont {Caneva},\ and\ \citenamefont
  {Chang}}]{PhysRevX.6.031017}%
  \BibitemOpen
  \bibfield  {author} {\bibinfo {author} {\bibfnamefont {J.~S.}\ \bibnamefont
  {Douglas}}, \bibinfo {author} {\bibfnamefont {T.}~\bibnamefont {Caneva}}, \
  and\ \bibinfo {author} {\bibfnamefont {D.~E.}\ \bibnamefont {Chang}},\ }\href
  {\doibase 10.1103/PhysRevX.6.031017} {\bibfield  {journal} {\bibinfo
  {journal} {Phys. Rev. X}\ }\textbf {\bibinfo {volume} {6}},\ \bibinfo {pages}
  {031017} (\bibinfo {year} {2016})}\BibitemShut {NoStop}%
\bibitem [{\citenamefont {Gonz\'alez-Tudela}\ \emph {et~al.}(2017)\citenamefont
  {Gonz\'alez-Tudela}, \citenamefont {Paulisch}, \citenamefont {Kimble},\ and\
  \citenamefont {Cirac}}]{PhysRevLett.118.213601}%
  \BibitemOpen
  \bibfield  {author} {\bibinfo {author} {\bibfnamefont {A.}~\bibnamefont
  {Gonz\'alez-Tudela}}, \bibinfo {author} {\bibfnamefont {V.}~\bibnamefont
  {Paulisch}}, \bibinfo {author} {\bibfnamefont {H.~J.}\ \bibnamefont
  {Kimble}}, \ and\ \bibinfo {author} {\bibfnamefont {J.~I.}\ \bibnamefont
  {Cirac}},\ }\href {\doibase 10.1103/PhysRevLett.118.213601} {\bibfield
  {journal} {\bibinfo  {journal} {Phys. Rev. Lett.}\ }\textbf {\bibinfo
  {volume} {118}},\ \bibinfo {pages} {213601} (\bibinfo {year}
  {2017})}\BibitemShut {NoStop}%
\bibitem [{\citenamefont {Chang}\ \emph {et~al.}(2018)\citenamefont {Chang},
  \citenamefont {Douglas}, \citenamefont {Gonz\'{a}lez-Tudela}, \citenamefont
  {Hung},\ and\ \citenamefont {Kimble}}]{Chang2018Aug}%
  \BibitemOpen
  \bibfield  {author} {\bibinfo {author} {\bibfnamefont {D.~E.}\ \bibnamefont
  {Chang}}, \bibinfo {author} {\bibfnamefont {J.~S.}\ \bibnamefont {Douglas}},
  \bibinfo {author} {\bibfnamefont {A.}~\bibnamefont {Gonz\'{a}lez-Tudela}},
  \bibinfo {author} {\bibfnamefont {C.-L.}\ \bibnamefont {Hung}}, \ and\
  \bibinfo {author} {\bibfnamefont {H.~J.}\ \bibnamefont {Kimble}},\ }\href
  {\doibase 10.1103/RevModPhys.90.031002} {\bibfield  {journal} {\bibinfo
  {journal} {Rev. Mod. Phys.}\ }\textbf {\bibinfo {volume} {90}},\ \bibinfo
  {pages} {031002} (\bibinfo {year} {2018})}\BibitemShut {NoStop}%
\bibitem [{\citenamefont {Bienias}\ \emph {et~al.}(2014)\citenamefont
  {Bienias}, \citenamefont {Choi}, \citenamefont {Firstenberg}, \citenamefont
  {Maghrebi}, \citenamefont {Gullans}, \citenamefont {Lukin}, \citenamefont
  {Gorshkov},\ and\ \citenamefont {B\"uchler}}]{PhysRevA.90.053804}%
  \BibitemOpen
  \bibfield  {author} {\bibinfo {author} {\bibfnamefont {P.}~\bibnamefont
  {Bienias}}, \bibinfo {author} {\bibfnamefont {S.}~\bibnamefont {Choi}},
  \bibinfo {author} {\bibfnamefont {O.}~\bibnamefont {Firstenberg}}, \bibinfo
  {author} {\bibfnamefont {M.~F.}\ \bibnamefont {Maghrebi}}, \bibinfo {author}
  {\bibfnamefont {M.}~\bibnamefont {Gullans}}, \bibinfo {author} {\bibfnamefont
  {M.~D.}\ \bibnamefont {Lukin}}, \bibinfo {author} {\bibfnamefont {A.~V.}\
  \bibnamefont {Gorshkov}}, \ and\ \bibinfo {author} {\bibfnamefont {H.~P.}\
  \bibnamefont {B\"uchler}},\ }\href {\doibase 10.1103/PhysRevA.90.053804}
  {\bibfield  {journal} {\bibinfo  {journal} {Phys. Rev. A}\ }\textbf {\bibinfo
  {volume} {90}},\ \bibinfo {pages} {053804} (\bibinfo {year}
  {2014})}\BibitemShut {NoStop}%
\bibitem [{\citenamefont {Maghrebi}\ \emph {et~al.}(2015)\citenamefont
  {Maghrebi}, \citenamefont {Gullans}, \citenamefont {Bienias}, \citenamefont
  {Choi}, \citenamefont {Martin}, \citenamefont {Firstenberg}, \citenamefont
  {Lukin}, \citenamefont {B\"uchler},\ and\ \citenamefont
  {Gorshkov}}]{PhysRevLett.115.123601}%
  \BibitemOpen
  \bibfield  {author} {\bibinfo {author} {\bibfnamefont {M.~F.}\ \bibnamefont
  {Maghrebi}}, \bibinfo {author} {\bibfnamefont {M.~J.}\ \bibnamefont
  {Gullans}}, \bibinfo {author} {\bibfnamefont {P.}~\bibnamefont {Bienias}},
  \bibinfo {author} {\bibfnamefont {S.}~\bibnamefont {Choi}}, \bibinfo {author}
  {\bibfnamefont {I.}~\bibnamefont {Martin}}, \bibinfo {author} {\bibfnamefont
  {O.}~\bibnamefont {Firstenberg}}, \bibinfo {author} {\bibfnamefont {M.~D.}\
  \bibnamefont {Lukin}}, \bibinfo {author} {\bibfnamefont {H.~P.}\ \bibnamefont
  {B\"uchler}}, \ and\ \bibinfo {author} {\bibfnamefont {A.~V.}\ \bibnamefont
  {Gorshkov}},\ }\href {\doibase 10.1103/PhysRevLett.115.123601} {\bibfield
  {journal} {\bibinfo  {journal} {Phys. Rev. Lett.}\ }\textbf {\bibinfo
  {volume} {115}},\ \bibinfo {pages} {123601} (\bibinfo {year}
  {2015})}\BibitemShut {NoStop}%
\bibitem [{\citenamefont {Liao}\ and\ \citenamefont {Law}(2010)}]{Liao2010Nov}%
  \BibitemOpen
  \bibfield  {author} {\bibinfo {author} {\bibfnamefont {J.-Q.}\ \bibnamefont
  {Liao}}\ and\ \bibinfo {author} {\bibfnamefont {C.~K.}\ \bibnamefont {Law}},\
  }\href {\doibase 10.1103/PhysRevA.82.053836} {\bibfield  {journal} {\bibinfo
  {journal} {Phys. Rev. A}\ }\textbf {\bibinfo {volume} {82}},\ \bibinfo
  {pages} {053836} (\bibinfo {year} {2010})}\BibitemShut {NoStop}%
\bibitem [{\citenamefont {Liao}\ and\ \citenamefont
  {Law}(2013)}]{liao2013Correlated}%
  \BibitemOpen
  \bibfield  {author} {\bibinfo {author} {\bibfnamefont {J.-Q.}\ \bibnamefont
  {Liao}}\ and\ \bibinfo {author} {\bibfnamefont {C.~K.}\ \bibnamefont {Law}},\
  }\href {\doibase 10.1103/PhysRevA.87.043809} {\bibfield  {journal} {\bibinfo
  {journal} {Phys. Rev. A}\ }\textbf {\bibinfo {volume} {87}},\ \bibinfo
  {pages} {043809} (\bibinfo {year} {2013})}\BibitemShut {NoStop}%
\bibitem [{\citenamefont {Qin}\ \emph {et~al.}(2019)\citenamefont {Qin},
  \citenamefont {Macr\`{\i}}, \citenamefont {Miranowicz}, \citenamefont
  {Savasta},\ and\ \citenamefont {Nori}}]{Qin2019Dec}%
  \BibitemOpen
  \bibfield  {author} {\bibinfo {author} {\bibfnamefont {W.}~\bibnamefont
  {Qin}}, \bibinfo {author} {\bibfnamefont {V.}~\bibnamefont {Macr\`{\i}}},
  \bibinfo {author} {\bibfnamefont {A.}~\bibnamefont {Miranowicz}}, \bibinfo
  {author} {\bibfnamefont {S.}~\bibnamefont {Savasta}}, \ and\ \bibinfo
  {author} {\bibfnamefont {F.}~\bibnamefont {Nori}},\ }\href {\doibase
  10.1103/PhysRevA.100.062501} {\bibfield  {journal} {\bibinfo  {journal}
  {Phys. Rev. A}\ }\textbf {\bibinfo {volume} {100}},\ \bibinfo {pages}
  {062501} (\bibinfo {year} {2019})}\BibitemShut {NoStop}%
\bibitem [{\citenamefont {Haroche}\ and\ \citenamefont
  {Raimond}(2006)}]{Haroche2006Oct}%
  \BibitemOpen
  \bibfield  {author} {\bibinfo {author} {\bibfnamefont {S.}~\bibnamefont
  {Haroche}}\ and\ \bibinfo {author} {\bibfnamefont {J.-M.}\ \bibnamefont
  {Raimond}},\ }\href@noop {} {\emph {\bibinfo {title} {{Exploring the Quantum:
  Atoms, Cavities, and Photons}}}}\ (\bibinfo  {publisher} {Oxford Univ.
  Press},\ \bibinfo {address} {Oxford, UK},\ \bibinfo {year}
  {2006})\BibitemShut {NoStop}%
\bibitem [{\citenamefont {Chang}\ \emph {et~al.}(2016)\citenamefont {Chang},
  \citenamefont {Gonz\'alez-Tudela}, \citenamefont {S\'anchez Mu\~noz},
  \citenamefont {Navarrete-Benlloch},\ and\ \citenamefont
  {Shi}}]{PhysRevLett.117.203602}%
  \BibitemOpen
  \bibfield  {author} {\bibinfo {author} {\bibfnamefont {Y.}~\bibnamefont
  {Chang}}, \bibinfo {author} {\bibfnamefont {A.}~\bibnamefont
  {Gonz\'alez-Tudela}}, \bibinfo {author} {\bibfnamefont {C.}~\bibnamefont
  {S\'anchez Mu\~noz}}, \bibinfo {author} {\bibfnamefont {C.}~\bibnamefont
  {Navarrete-Benlloch}}, \ and\ \bibinfo {author} {\bibfnamefont
  {T.}~\bibnamefont {Shi}},\ }\href {\doibase 10.1103/PhysRevLett.117.203602}
  {\bibfield  {journal} {\bibinfo  {journal} {Phys. Rev. Lett.}\ }\textbf
  {\bibinfo {volume} {117}},\ \bibinfo {pages} {203602} (\bibinfo {year}
  {2016})}\BibitemShut {NoStop}%
\bibitem [{\citenamefont {Dong}\ and\ \citenamefont
  {Li}(2019)}]{PhysRevA.100.043825}%
  \BibitemOpen
  \bibfield  {author} {\bibinfo {author} {\bibfnamefont {X.-L.}\ \bibnamefont
  {Dong}}\ and\ \bibinfo {author} {\bibfnamefont {P.-B.}\ \bibnamefont {Li}},\
  }\href {\doibase 10.1103/PhysRevA.100.043825} {\bibfield  {journal} {\bibinfo
   {journal} {Phys. Rev. A}\ }\textbf {\bibinfo {volume} {100}},\ \bibinfo
  {pages} {043825} (\bibinfo {year} {2019})}\BibitemShut {NoStop}%
\bibitem [{\citenamefont {M\"{u}ller}\ \emph {et~al.}(2014)\citenamefont
  {M\"{u}ller}, \citenamefont {Bounouar}, \citenamefont {J\"{o}ns},
  \citenamefont {Gl\"{a}ssl},\ and\ \citenamefont {Michler}}]{Muller2014Mar}%
  \BibitemOpen
  \bibfield  {author} {\bibinfo {author} {\bibfnamefont {M.}~\bibnamefont
  {M\"{u}ller}}, \bibinfo {author} {\bibfnamefont {S.}~\bibnamefont
  {Bounouar}}, \bibinfo {author} {\bibfnamefont {K.~D.}\ \bibnamefont
  {J\"{o}ns}}, \bibinfo {author} {\bibfnamefont {M.}~\bibnamefont
  {Gl\"{a}ssl}}, \ and\ \bibinfo {author} {\bibfnamefont {P.}~\bibnamefont
  {Michler}},\ }\href {\doibase 10.1038/nphoton.2013.377} {\bibfield  {journal}
  {\bibinfo  {journal} {Nat. Photonics}\ }\textbf {\bibinfo {volume} {8}},\
  \bibinfo {pages} {224} (\bibinfo {year} {2014})}\BibitemShut {NoStop}%
\bibitem [{\citenamefont {Hargart}\ \emph {et~al.}(2016)\citenamefont
  {Hargart}, \citenamefont {M\"{u}ller}, \citenamefont {Roy-Choudhury},
  \citenamefont {Portalupi}, \citenamefont {Schneider}, \citenamefont
  {H\"{o}fling}, \citenamefont {Kamp}, \citenamefont {Hughes},\ and\
  \citenamefont {Michler}}]{Hargart2016Mar}%
  \BibitemOpen
  \bibfield  {author} {\bibinfo {author} {\bibfnamefont {F.}~\bibnamefont
  {Hargart}}, \bibinfo {author} {\bibfnamefont {M.}~\bibnamefont {M\"{u}ller}},
  \bibinfo {author} {\bibfnamefont {K.}~\bibnamefont {Roy-Choudhury}}, \bibinfo
  {author} {\bibfnamefont {S.~L.}\ \bibnamefont {Portalupi}}, \bibinfo {author}
  {\bibfnamefont {C.}~\bibnamefont {Schneider}}, \bibinfo {author}
  {\bibfnamefont {S.}~\bibnamefont {H\"{o}fling}}, \bibinfo {author}
  {\bibfnamefont {M.}~\bibnamefont {Kamp}}, \bibinfo {author} {\bibfnamefont
  {S.}~\bibnamefont {Hughes}}, \ and\ \bibinfo {author} {\bibfnamefont
  {P.}~\bibnamefont {Michler}},\ }\href {\doibase 10.1103/PhysRevB.93.115308}
  {\bibfield  {journal} {\bibinfo  {journal} {Phys. Rev. B}\ }\textbf {\bibinfo
  {volume} {93}},\ \bibinfo {pages} {115308} (\bibinfo {year}
  {2016})}\BibitemShut {NoStop}%
\bibitem [{\citenamefont {Koshino}\ \emph {et~al.}(2013)\citenamefont
  {Koshino}, \citenamefont {Inomata}, \citenamefont {Yamamoto},\ and\
  \citenamefont {Nakamura}}]{Koshino2013Oct}%
  \BibitemOpen
  \bibfield  {author} {\bibinfo {author} {\bibfnamefont {K.}~\bibnamefont
  {Koshino}}, \bibinfo {author} {\bibfnamefont {K.}~\bibnamefont {Inomata}},
  \bibinfo {author} {\bibfnamefont {T.}~\bibnamefont {Yamamoto}}, \ and\
  \bibinfo {author} {\bibfnamefont {Y.}~\bibnamefont {Nakamura}},\ }\href
  {\doibase 10.1103/PhysRevLett.111.153601} {\bibfield  {journal} {\bibinfo
  {journal} {Phys. Rev. Lett.}\ }\textbf {\bibinfo {volume} {111}},\ \bibinfo
  {pages} {153601} (\bibinfo {year} {2013})}\BibitemShut {NoStop}%
\bibitem [{\citenamefont {Dousse}\ \emph {et~al.}(2010)\citenamefont {Dousse},
  \citenamefont {Suffczy\'{n}ski}, \citenamefont {Beveratos}, \citenamefont
  {Krebs}, \citenamefont {Lema\^{\i}tre}, \citenamefont {Sagnes}, \citenamefont
  {Bloch}, \citenamefont {Voisin},\ and\ \citenamefont
  {Senellart}}]{Dousse2010Jul}%
  \BibitemOpen
  \bibfield  {author} {\bibinfo {author} {\bibfnamefont {A.}~\bibnamefont
  {Dousse}}, \bibinfo {author} {\bibfnamefont {J.}~\bibnamefont
  {Suffczy\'{n}ski}}, \bibinfo {author} {\bibfnamefont {A.}~\bibnamefont
  {Beveratos}}, \bibinfo {author} {\bibfnamefont {O.}~\bibnamefont {Krebs}},
  \bibinfo {author} {\bibfnamefont {A.}~\bibnamefont {Lema\^{\i}tre}}, \bibinfo
  {author} {\bibfnamefont {I.}~\bibnamefont {Sagnes}}, \bibinfo {author}
  {\bibfnamefont {J.}~\bibnamefont {Bloch}}, \bibinfo {author} {\bibfnamefont
  {P.}~\bibnamefont {Voisin}}, \ and\ \bibinfo {author} {\bibfnamefont
  {P.}~\bibnamefont {Senellart}},\ }\href {\doibase 10.1038/nature09148}
  {\bibfield  {journal} {\bibinfo  {journal} {Nature}\ }\textbf {\bibinfo
  {volume} {466}},\ \bibinfo {pages} {217} (\bibinfo {year}
  {2010})}\BibitemShut {NoStop}%
\bibitem [{\citenamefont {Ota}\ \emph {et~al.}(2011)\citenamefont {Ota},
  \citenamefont {Iwamoto}, \citenamefont {Kumagai},\ and\ \citenamefont
  {Arakawa}}]{Ota2011Nov}%
  \BibitemOpen
  \bibfield  {author} {\bibinfo {author} {\bibfnamefont {Y.}~\bibnamefont
  {Ota}}, \bibinfo {author} {\bibfnamefont {S.}~\bibnamefont {Iwamoto}},
  \bibinfo {author} {\bibfnamefont {N.}~\bibnamefont {Kumagai}}, \ and\
  \bibinfo {author} {\bibfnamefont {Y.}~\bibnamefont {Arakawa}},\ }\href
  {\doibase 10.1103/PhysRevLett.107.233602} {\bibfield  {journal} {\bibinfo
  {journal} {Phys. Rev. Lett.}\ }\textbf {\bibinfo {volume} {107}},\ \bibinfo
  {pages} {233602} (\bibinfo {year} {2011})}\BibitemShut {NoStop}%
\bibitem [{\citenamefont {S\'{a}nchez-Burillo}\ \emph
  {et~al.}(2016)\citenamefont {S\'{a}nchez-Burillo}, \citenamefont
  {Mart\'{\i}n-Moreno}, \citenamefont {Garc\'{\i}a-Ripoll},\ and\ \citenamefont
  {Zueco}}]{Sanchez-Burillo2016Nov}%
  \BibitemOpen
  \bibfield  {author} {\bibinfo {author} {\bibfnamefont {E.}~\bibnamefont
  {S\'{a}nchez-Burillo}}, \bibinfo {author} {\bibfnamefont {L.}~\bibnamefont
  {Mart\'{\i}n-Moreno}}, \bibinfo {author} {\bibfnamefont {J.~J.}\ \bibnamefont
  {Garc\'{\i}a-Ripoll}}, \ and\ \bibinfo {author} {\bibfnamefont
  {D.}~\bibnamefont {Zueco}},\ }\href {\doibase 10.1103/PhysRevA.94.053814}
  {\bibfield  {journal} {\bibinfo  {journal} {Phys. Rev. A}\ }\textbf {\bibinfo
  {volume} {94}},\ \bibinfo {pages} {053814} (\bibinfo {year}
  {2016})}\BibitemShut {NoStop}%
\bibitem [{\citenamefont {Imamo{\=g}lu}\ \emph {et~al.}(1997)\citenamefont
  {Imamo{\=g}lu}, \citenamefont {Schmidt}, \citenamefont {Woods},\ and\
  \citenamefont {Deutsch}}]{xn--Imamolu-4s3c1997Aug}%
  \BibitemOpen
  \bibfield  {author} {\bibinfo {author} {\bibfnamefont {A.}~\bibnamefont
  {Imamo{\=g}lu}}, \bibinfo {author} {\bibfnamefont {H.}~\bibnamefont
  {Schmidt}}, \bibinfo {author} {\bibfnamefont {G.}~\bibnamefont {Woods}}, \
  and\ \bibinfo {author} {\bibfnamefont {M.}~\bibnamefont {Deutsch}},\ }\href
  {\doibase 10.1103/PhysRevLett.79.1467} {\bibfield  {journal} {\bibinfo
  {journal} {Phys. Rev. Lett.}\ }\textbf {\bibinfo {volume} {79}},\ \bibinfo
  {pages} {1467} (\bibinfo {year} {1997})}\BibitemShut {NoStop}%
\bibitem [{\citenamefont {Birnbaum}\ \emph {et~al.}(2005)\citenamefont
  {Birnbaum}, \citenamefont {Boca}, \citenamefont {Miller}, \citenamefont
  {Boozer}, \citenamefont {Northup},\ and\ \citenamefont
  {Kimble}}]{Birnbaum2005Jul}%
  \BibitemOpen
  \bibfield  {author} {\bibinfo {author} {\bibfnamefont {K.~M.}\ \bibnamefont
  {Birnbaum}}, \bibinfo {author} {\bibfnamefont {A.}~\bibnamefont {Boca}},
  \bibinfo {author} {\bibfnamefont {R.}~\bibnamefont {Miller}}, \bibinfo
  {author} {\bibfnamefont {A.~D.}\ \bibnamefont {Boozer}}, \bibinfo {author}
  {\bibfnamefont {T.~E.}\ \bibnamefont {Northup}}, \ and\ \bibinfo {author}
  {\bibfnamefont {H.~J.}\ \bibnamefont {Kimble}},\ }\href {\doibase
  10.1038/nature03804} {\bibfield  {journal} {\bibinfo  {journal} {Nature}\
  }\textbf {\bibinfo {volume} {436}},\ \bibinfo {pages} {87} (\bibinfo {year}
  {2005})}\BibitemShut {NoStop}%
\bibitem [{\citenamefont {Liew}\ and\ \citenamefont
  {Savona}(2010)}]{Liew2010May}%
  \BibitemOpen
  \bibfield  {author} {\bibinfo {author} {\bibfnamefont {T.~C.~H.}\
  \bibnamefont {Liew}}\ and\ \bibinfo {author} {\bibfnamefont {V.}~\bibnamefont
  {Savona}},\ }\href {\doibase 10.1103/PhysRevLett.104.183601} {\bibfield
  {journal} {\bibinfo  {journal} {Phys. Rev. Lett.}\ }\textbf {\bibinfo
  {volume} {104}},\ \bibinfo {pages} {183601} (\bibinfo {year}
  {2010})}\BibitemShut {NoStop}%
\bibitem [{\citenamefont {Rabl}(2011)}]{Rabl2011Aug}%
  \BibitemOpen
  \bibfield  {author} {\bibinfo {author} {\bibfnamefont {P.}~\bibnamefont
  {Rabl}},\ }\href {\doibase 10.1103/PhysRevLett.107.063601} {\bibfield
  {journal} {\bibinfo  {journal} {Phys. Rev. Lett.}\ }\textbf {\bibinfo
  {volume} {107}},\ \bibinfo {pages} {063601} (\bibinfo {year}
  {2011})}\BibitemShut {NoStop}%
\bibitem [{\citenamefont {Ridolfo}\ \emph {et~al.}(2012)\citenamefont
  {Ridolfo}, \citenamefont {Leib}, \citenamefont {Savasta},\ and\ \citenamefont
  {Hartmann}}]{Ridolfo2012Nov}%
  \BibitemOpen
  \bibfield  {author} {\bibinfo {author} {\bibfnamefont {A.}~\bibnamefont
  {Ridolfo}}, \bibinfo {author} {\bibfnamefont {M.}~\bibnamefont {Leib}},
  \bibinfo {author} {\bibfnamefont {S.}~\bibnamefont {Savasta}}, \ and\
  \bibinfo {author} {\bibfnamefont {M.~J.}\ \bibnamefont {Hartmann}},\ }\href
  {\doibase 10.1103/PhysRevLett.109.193602} {\bibfield  {journal} {\bibinfo
  {journal} {Phys. Rev. Lett.}\ }\textbf {\bibinfo {volume} {109}},\ \bibinfo
  {pages} {193602} (\bibinfo {year} {2012})}\BibitemShut {NoStop}%
\bibitem [{\citenamefont {Liao}\ and\ \citenamefont
  {Nori}(2013)}]{Liao2013Aug}%
  \BibitemOpen
  \bibfield  {author} {\bibinfo {author} {\bibfnamefont {J.-Q.}\ \bibnamefont
  {Liao}}\ and\ \bibinfo {author} {\bibfnamefont {F.}~\bibnamefont {Nori}},\
  }\href {\doibase 10.1103/PhysRevA.88.023853} {\bibfield  {journal} {\bibinfo
  {journal} {Phys. Rev. A}\ }\textbf {\bibinfo {volume} {88}},\ \bibinfo
  {pages} {023853} (\bibinfo {year} {2013})}\BibitemShut {NoStop}%
\bibitem [{\citenamefont {Miranowicz}\ \emph {et~al.}(2013)\citenamefont
  {Miranowicz}, \citenamefont {Paprzycka}, \citenamefont {Liu}, \citenamefont
  {Bajer},\ and\ \citenamefont {Nori}}]{Miranowicz2013Feb}%
  \BibitemOpen
  \bibfield  {author} {\bibinfo {author} {\bibfnamefont {A.}~\bibnamefont
  {Miranowicz}}, \bibinfo {author} {\bibfnamefont {M.}~\bibnamefont
  {Paprzycka}}, \bibinfo {author} {\bibfnamefont {Y.-X.}\ \bibnamefont {Liu}},
  \bibinfo {author} {\bibfnamefont {J.}~\bibnamefont {Bajer}}, \ and\ \bibinfo
  {author} {\bibfnamefont {F.}~\bibnamefont {Nori}},\ }\href {\doibase
  10.1103/PhysRevA.87.023809} {\bibfield  {journal} {\bibinfo  {journal} {Phys.
  Rev. A}\ }\textbf {\bibinfo {volume} {87}},\ \bibinfo {pages} {023809}
  (\bibinfo {year} {2013})}\BibitemShut {NoStop}%
\bibitem [{\citenamefont {Snijders}\ \emph {et~al.}(2018)\citenamefont
  {Snijders}, \citenamefont {Frey}, \citenamefont {Norman}, \citenamefont
  {Flayac}, \citenamefont {Savona}, \citenamefont {Gossard}, \citenamefont
  {Bowers}, \citenamefont {van Exter}, \citenamefont {Bouwmeester},\ and\
  \citenamefont {L{\ifmmode\ddot{o}\else\"{o}\fi}ffler}}]{Snijders2018Jul}%
  \BibitemOpen
  \bibfield  {author} {\bibinfo {author} {\bibfnamefont {H.~J.}\ \bibnamefont
  {Snijders}}, \bibinfo {author} {\bibfnamefont {J.~A.}\ \bibnamefont {Frey}},
  \bibinfo {author} {\bibfnamefont {J.}~\bibnamefont {Norman}}, \bibinfo
  {author} {\bibfnamefont {H.}~\bibnamefont {Flayac}}, \bibinfo {author}
  {\bibfnamefont {V.}~\bibnamefont {Savona}}, \bibinfo {author} {\bibfnamefont
  {A.~C.}\ \bibnamefont {Gossard}}, \bibinfo {author} {\bibfnamefont {J.~E.}\
  \bibnamefont {Bowers}}, \bibinfo {author} {\bibfnamefont {M.~P.}\
  \bibnamefont {van Exter}}, \bibinfo {author} {\bibfnamefont {D.}~\bibnamefont
  {Bouwmeester}}, \ and\ \bibinfo {author} {\bibfnamefont {W.}~\bibnamefont
  {L{\ifmmode\ddot{o}\else\"{o}\fi}ffler}},\ }\href {\doibase
  10.1103/PhysRevLett.121.043601} {\bibfield  {journal} {\bibinfo  {journal}
  {Phys. Rev. Lett.}\ }\textbf {\bibinfo {volume} {121}},\ \bibinfo {pages}
  {043601} (\bibinfo {year} {2018})}\BibitemShut {NoStop}%
\bibitem [{\citenamefont {Vaneph}\ \emph {et~al.}(2018)\citenamefont {Vaneph},
  \citenamefont {Morvan}, \citenamefont {Aiello}, \citenamefont {F\'{e}chant},
  \citenamefont {Aprili}, \citenamefont {Gabelli},\ and\ \citenamefont
  {Est\`{e}ve}}]{Vaneph2018Jul}%
  \BibitemOpen
  \bibfield  {author} {\bibinfo {author} {\bibfnamefont {C.}~\bibnamefont
  {Vaneph}}, \bibinfo {author} {\bibfnamefont {A.}~\bibnamefont {Morvan}},
  \bibinfo {author} {\bibfnamefont {G.}~\bibnamefont {Aiello}}, \bibinfo
  {author} {\bibfnamefont {M.}~\bibnamefont {F\'{e}chant}}, \bibinfo {author}
  {\bibfnamefont {M.}~\bibnamefont {Aprili}}, \bibinfo {author} {\bibfnamefont
  {J.}~\bibnamefont {Gabelli}}, \ and\ \bibinfo {author} {\bibfnamefont
  {J.}~\bibnamefont {Est\`{e}ve}},\ }\href {\doibase
  10.1103/PhysRevLett.121.043602} {\bibfield  {journal} {\bibinfo  {journal}
  {Phys. Rev. Lett.}\ }\textbf {\bibinfo {volume} {121}},\ \bibinfo {pages}
  {043602} (\bibinfo {year} {2018})}\BibitemShut {NoStop}%
\bibitem [{\citenamefont {Huang}\ \emph {et~al.}(2018)\citenamefont {Huang},
  \citenamefont {Miranowicz}, \citenamefont {Liao}, \citenamefont {Nori},\ and\
  \citenamefont {Jing}}]{Huang2018Oct}%
  \BibitemOpen
  \bibfield  {author} {\bibinfo {author} {\bibfnamefont {R.}~\bibnamefont
  {Huang}}, \bibinfo {author} {\bibfnamefont {A.}~\bibnamefont {Miranowicz}},
  \bibinfo {author} {\bibfnamefont {J.-Q.}\ \bibnamefont {Liao}}, \bibinfo
  {author} {\bibfnamefont {F.}~\bibnamefont {Nori}}, \ and\ \bibinfo {author}
  {\bibfnamefont {H.}~\bibnamefont {Jing}},\ }\href {\doibase
  10.1103/PhysRevLett.121.153601} {\bibfield  {journal} {\bibinfo  {journal}
  {Phys. Rev. Lett.}\ }\textbf {\bibinfo {volume} {121}},\ \bibinfo {pages}
  {153601} (\bibinfo {year} {2018})}\BibitemShut {NoStop}%
\bibitem [{\citenamefont {Zou}\ \emph {et~al.}(2019)\citenamefont {Zou},
  \citenamefont {Fan}, \citenamefont {Huang},\ and\ \citenamefont
  {Liao}}]{Zou2019Apr}%
  \BibitemOpen
  \bibfield  {author} {\bibinfo {author} {\bibfnamefont {F.}~\bibnamefont
  {Zou}}, \bibinfo {author} {\bibfnamefont {L.-B.}\ \bibnamefont {Fan}},
  \bibinfo {author} {\bibfnamefont {J.-F.}\ \bibnamefont {Huang}}, \ and\
  \bibinfo {author} {\bibfnamefont {J.-Q.}\ \bibnamefont {Liao}},\ }\href
  {\doibase 10.1103/PhysRevA.99.043837} {\bibfield  {journal} {\bibinfo
  {journal} {Phys. Rev. A}\ }\textbf {\bibinfo {volume} {99}},\ \bibinfo
  {pages} {043837} (\bibinfo {year} {2019})}\BibitemShut {NoStop}%
\bibitem [{\citenamefont {Xu}\ \emph {et~al.}(2020)\citenamefont {Xu},
  \citenamefont {Zhao}, \citenamefont {Wang}, \citenamefont {Jing},\ and\
  \citenamefont {Chen}}]{Xu2020Feb}%
  \BibitemOpen
  \bibfield  {author} {\bibinfo {author} {\bibfnamefont {X.-W.}\ \bibnamefont
  {Xu}}, \bibinfo {author} {\bibfnamefont {Y.-J.}\ \bibnamefont {Zhao}},
  \bibinfo {author} {\bibfnamefont {H.}~\bibnamefont {Wang}}, \bibinfo {author}
  {\bibfnamefont {H.}~\bibnamefont {Jing}}, \ and\ \bibinfo {author}
  {\bibfnamefont {A.}~\bibnamefont {Chen}},\ }\href {\doibase
  10.1364/PRJ.8.000143} {\bibfield  {journal} {\bibinfo  {journal} {Photonics
  Res.}\ }\textbf {\bibinfo {volume} {8}},\ \bibinfo {pages} {143} (\bibinfo
  {year} {2020})}\BibitemShut {NoStop}%
\bibitem [{\citenamefont {Zou}\ \emph {et~al.}(2020)\citenamefont {Zou},
  \citenamefont {Zhang}, \citenamefont {Xu}, \citenamefont {Huang},\ and\
  \citenamefont {Liao}}]{Zou2020Nov}%
  \BibitemOpen
  \bibfield  {author} {\bibinfo {author} {\bibfnamefont {F.}~\bibnamefont
  {Zou}}, \bibinfo {author} {\bibfnamefont {X.-Y.}\ \bibnamefont {Zhang}},
  \bibinfo {author} {\bibfnamefont {X.-W.}\ \bibnamefont {Xu}}, \bibinfo
  {author} {\bibfnamefont {J.-F.}\ \bibnamefont {Huang}}, \ and\ \bibinfo
  {author} {\bibfnamefont {J.-Q.}\ \bibnamefont {Liao}},\ }\href {\doibase
  10.1103/PhysRevA.102.053710} {\bibfield  {journal} {\bibinfo  {journal}
  {Phys. Rev. A}\ }\textbf {\bibinfo {volume} {102}},\ \bibinfo {pages}
  {053710} (\bibinfo {year} {2020})}\BibitemShut {NoStop}%
\bibitem [{\citenamefont {Ren}\ \emph {et~al.}(2021)\citenamefont {Ren},
  \citenamefont {Duan}, \citenamefont {Xie}, \citenamefont {Shao},\ and\
  \citenamefont {Duan}}]{Ren2021May}%
  \BibitemOpen
  \bibfield  {author} {\bibinfo {author} {\bibfnamefont {Y.}~\bibnamefont
  {Ren}}, \bibinfo {author} {\bibfnamefont {S.}~\bibnamefont {Duan}}, \bibinfo
  {author} {\bibfnamefont {W.}~\bibnamefont {Xie}}, \bibinfo {author}
  {\bibfnamefont {Y.}~\bibnamefont {Shao}}, \ and\ \bibinfo {author}
  {\bibfnamefont {Z.}~\bibnamefont {Duan}},\ }\href {\doibase
  10.1103/PhysRevA.103.053710} {\bibfield  {journal} {\bibinfo  {journal}
  {Phys. Rev. A}\ }\textbf {\bibinfo {volume} {103}},\ \bibinfo {pages}
  {053710} (\bibinfo {year} {2021})}\BibitemShut {NoStop}%
\bibitem [{\citenamefont {Deng}\ \emph
  {et~al.}(2021{\natexlab{a}})\citenamefont {Deng}, \citenamefont {Zou},
  \citenamefont {Huang},\ and\ \citenamefont {Liao}}]{Deng2021Sep}%
  \BibitemOpen
  \bibfield  {author} {\bibinfo {author} {\bibfnamefont {H.}~\bibnamefont
  {Deng}}, \bibinfo {author} {\bibfnamefont {F.}~\bibnamefont {Zou}}, \bibinfo
  {author} {\bibfnamefont {J.-F.}\ \bibnamefont {Huang}}, \ and\ \bibinfo
  {author} {\bibfnamefont {J.-Q.}\ \bibnamefont {Liao}},\ }\href {\doibase
  10.1103/PhysRevA.104.033706} {\bibfield  {journal} {\bibinfo  {journal}
  {Phys. Rev. A}\ }\textbf {\bibinfo {volume} {104}},\ \bibinfo {pages}
  {033706} (\bibinfo {year} {2021}{\natexlab{a}})}\BibitemShut {NoStop}%
\bibitem [{\citenamefont {Strekalov}(2014)}]{Strekalov2014Jul}%
  \BibitemOpen
  \bibfield  {author} {\bibinfo {author} {\bibfnamefont {D.~V.}\ \bibnamefont
  {Strekalov}},\ }\href {\doibase 10.1038/nphoton.2014.144} {\bibfield
  {journal} {\bibinfo  {journal} {Nat. Photonics}\ }\textbf {\bibinfo {volume}
  {8}},\ \bibinfo {pages} {500} (\bibinfo {year} {2014})}\BibitemShut {NoStop}%
\bibitem [{\citenamefont {Mu{\~{n}}oz}\ \emph {et~al.}(2018)\citenamefont
  {Mu{\~{n}}oz}, \citenamefont {Laussy}, \citenamefont {del Valle},
  \citenamefont {Tejedor},\ and\ \citenamefont
  {Gonz\'{a}lez-Tudela}}]{10.1364/OPTICA.5.000014}%
  \BibitemOpen
  \bibfield  {author} {\bibinfo {author} {\bibfnamefont {C.~S.}\ \bibnamefont
  {Mu{\~{n}}oz}}, \bibinfo {author} {\bibfnamefont {F.~P.}\ \bibnamefont
  {Laussy}}, \bibinfo {author} {\bibfnamefont {E.}~\bibnamefont {del Valle}},
  \bibinfo {author} {\bibfnamefont {C.}~\bibnamefont {Tejedor}}, \ and\
  \bibinfo {author} {\bibfnamefont {A.}~\bibnamefont {Gonz\'{a}lez-Tudela}},\
  }\href {\doibase 10.1364/OPTICA.5.000014} {\bibfield  {journal} {\bibinfo
  {journal} {Optica}\ }\textbf {\bibinfo {volume} {5}},\ \bibinfo {pages} {14}
  (\bibinfo {year} {2018})}\BibitemShut {NoStop}%
\bibitem [{\citenamefont {Bin}\ \emph {et~al.}(2020)\citenamefont {Bin},
  \citenamefont {L\"u}, \citenamefont {Laussy}, \citenamefont {Nori},\ and\
  \citenamefont {Wu}}]{PhysRevLett.124.053601}%
  \BibitemOpen
  \bibfield  {author} {\bibinfo {author} {\bibfnamefont {Q.}~\bibnamefont
  {Bin}}, \bibinfo {author} {\bibfnamefont {X.-Y.}\ \bibnamefont {L\"u}},
  \bibinfo {author} {\bibfnamefont {F.~P.}\ \bibnamefont {Laussy}}, \bibinfo
  {author} {\bibfnamefont {F.}~\bibnamefont {Nori}}, \ and\ \bibinfo {author}
  {\bibfnamefont {Y.}~\bibnamefont {Wu}},\ }\href {\doibase
  10.1103/PhysRevLett.124.053601} {\bibfield  {journal} {\bibinfo  {journal}
  {Phys. Rev. Lett.}\ }\textbf {\bibinfo {volume} {124}},\ \bibinfo {pages}
  {053601} (\bibinfo {year} {2020})}\BibitemShut {NoStop}%
\bibitem [{\citenamefont {Bin}\ \emph {et~al.}(2021)\citenamefont {Bin},
  \citenamefont {Wu},\ and\ \citenamefont
  {L\"u}}]{bin2020paritysymmetryprotected}%
  \BibitemOpen
  \bibfield  {author} {\bibinfo {author} {\bibfnamefont {Q.}~\bibnamefont
  {Bin}}, \bibinfo {author} {\bibfnamefont {Y.}~\bibnamefont {Wu}}, \ and\
  \bibinfo {author} {\bibfnamefont {X.-Y.}\ \bibnamefont {L\"u}},\ }\href
  {\doibase 10.1103/PhysRevLett.127.073602} {\bibfield  {journal} {\bibinfo
  {journal} {Phys. Rev. Lett.}\ }\textbf {\bibinfo {volume} {127}},\ \bibinfo
  {pages} {073602} (\bibinfo {year} {2021})}\BibitemShut {NoStop}%
\bibitem [{\citenamefont {Deng}\ \emph
  {et~al.}(2021{\natexlab{b}})\citenamefont {Deng}, \citenamefont {Shi},\ and\
  \citenamefont {Yi}}]{Deng:21}%
  \BibitemOpen
  \bibfield  {author} {\bibinfo {author} {\bibfnamefont {Y.~G.}\ \bibnamefont
  {Deng}}, \bibinfo {author} {\bibfnamefont {T.}~\bibnamefont {Shi}}, \ and\
  \bibinfo {author} {\bibfnamefont {S.}~\bibnamefont {Yi}},\ }\href {\doibase
  10.1364/PRJ.427062} {\bibfield  {journal} {\bibinfo  {journal} {Photon.
  Res.}\ }\textbf {\bibinfo {volume} {9}},\ \bibinfo {pages} {1289} (\bibinfo
  {year} {2021}{\natexlab{b}})}\BibitemShut {NoStop}%
\bibitem [{\citenamefont {Cosacchi}\ \emph {et~al.}(2021)\citenamefont
  {Cosacchi}, \citenamefont {Mielnik-Pyszczorski}, \citenamefont {Seidelmann},
  \citenamefont {Cygorek}, \citenamefont {Vagov}, \citenamefont {Reiter},\ and\
  \citenamefont {Axt}}]{Cosacchi2021Aug}%
  \BibitemOpen
  \bibfield  {author} {\bibinfo {author} {\bibfnamefont {M.}~\bibnamefont
  {Cosacchi}}, \bibinfo {author} {\bibfnamefont {A.}~\bibnamefont
  {Mielnik-Pyszczorski}}, \bibinfo {author} {\bibfnamefont {T.}~\bibnamefont
  {Seidelmann}}, \bibinfo {author} {\bibfnamefont {M.}~\bibnamefont {Cygorek}},
  \bibinfo {author} {\bibfnamefont {A.}~\bibnamefont {Vagov}}, \bibinfo
  {author} {\bibfnamefont {D.~E.}\ \bibnamefont {Reiter}}, \ and\ \bibinfo
  {author} {\bibfnamefont {V.~M.}\ \bibnamefont {Axt}},\ }\href
  {https://ui.adsabs.harvard.edu/abs/2021arXiv210803967C/abstract} {\bibfield
  {journal} {\bibinfo  {journal} {arXiv:2108.03967}\ } (\bibinfo {year}
  {2021})},\ \Eprint {http://arxiv.org/abs/2108.03967} {2108.03967}
  \BibitemShut {NoStop}%
\bibitem [{\citenamefont {D\'{\i}az-Camacho}\ \emph {et~al.}(2021)\citenamefont
  {D\'{\i}az-Camacho}, \citenamefont {Casalengua}, \citenamefont {Carre\~{n}o},
  \citenamefont {Khalid}, \citenamefont {Tejedor}, \citenamefont {del Valle},\
  and\ \citenamefont {Laussy}}]{Diaz-Camacho2021Sep}%
  \BibitemOpen
  \bibfield  {author} {\bibinfo {author} {\bibfnamefont {G.}~\bibnamefont
  {D\'{\i}az-Camacho}}, \bibinfo {author} {\bibfnamefont {E.~Z.}\ \bibnamefont
  {Casalengua}}, \bibinfo {author} {\bibfnamefont {J.~C.~L.}\ \bibnamefont
  {Carre\~{n}o}}, \bibinfo {author} {\bibfnamefont {S.}~\bibnamefont {Khalid}},
  \bibinfo {author} {\bibfnamefont {C.}~\bibnamefont {Tejedor}}, \bibinfo
  {author} {\bibfnamefont {E.}~\bibnamefont {del Valle}}, \ and\ \bibinfo
  {author} {\bibfnamefont {F.~P.}\ \bibnamefont {Laussy}},\ }\href
  {https://doi.org/10.48550/arXiv.2109.12049} {\bibfield  {journal} {\bibinfo
  {journal} {arXiv:2108.03967}\ } (\bibinfo {year} {2021})},\ \Eprint
  {http://arxiv.org/abs/2109.12049} {2109.12049} \BibitemShut {NoStop}%
\bibitem [{\citenamefont {Ma}\ \emph {et~al.}(2021)\citenamefont {Ma},
  \citenamefont {Li}, \citenamefont {Ren}, \citenamefont {Xie},\ and\
  \citenamefont {Li}}]{Ma2021Oct}%
  \BibitemOpen
  \bibfield  {author} {\bibinfo {author} {\bibfnamefont {S.-l.}\ \bibnamefont
  {Ma}}, \bibinfo {author} {\bibfnamefont {X.-k.}\ \bibnamefont {Li}}, \bibinfo
  {author} {\bibfnamefont {Y.-l.}\ \bibnamefont {Ren}}, \bibinfo {author}
  {\bibfnamefont {J.-k.}\ \bibnamefont {Xie}}, \ and\ \bibinfo {author}
  {\bibfnamefont {F.-l.}\ \bibnamefont {Li}},\ }\href {\doibase
  10.1103/PhysRevResearch.3.043020} {\bibfield  {journal} {\bibinfo  {journal}
  {Phys. Rev. Research}\ }\textbf {\bibinfo {volume} {3}},\ \bibinfo {pages}
  {043020} (\bibinfo {year} {2021})}\BibitemShut {NoStop}%
\bibitem [{\citenamefont {Zou}\ \emph {et~al.}(2022)\citenamefont {Zou},
  \citenamefont {Liao},\ and\ \citenamefont {Li}}]{Zou2021Dec}%
  \BibitemOpen
  \bibfield  {author} {\bibinfo {author} {\bibfnamefont {F.}~\bibnamefont
  {Zou}}, \bibinfo {author} {\bibfnamefont {J.-Q.}\ \bibnamefont {Liao}}, \
  and\ \bibinfo {author} {\bibfnamefont {Y.}~\bibnamefont {Li}},\ }\href
  {\doibase 10.1103/PhysRevA.105.053507} {\bibfield  {journal} {\bibinfo
  {journal} {Phys. Rev. A}\ }\textbf {\bibinfo {volume} {105}},\ \bibinfo
  {pages} {053507} (\bibinfo {year} {2022})}\BibitemShut {NoStop}%
\bibitem [{\citenamefont {M\'enard}\ \emph {et~al.}(2022)\citenamefont
  {M\'enard}, \citenamefont {Peugeot}, \citenamefont {Padurariu}, \citenamefont
  {Rolland}, \citenamefont {Kubala}, \citenamefont {Mukharsky}, \citenamefont
  {Iftikhar}, \citenamefont {Altimiras}, \citenamefont {Roche}, \citenamefont
  {le~Sueur}, \citenamefont {Joyez}, \citenamefont {Vion}, \citenamefont
  {Esteve}, \citenamefont {Ankerhold},\ and\ \citenamefont
  {Portier}}]{Menard2022Apr}%
  \BibitemOpen
  \bibfield  {author} {\bibinfo {author} {\bibfnamefont {G.~C.}\ \bibnamefont
  {M\'enard}}, \bibinfo {author} {\bibfnamefont {A.}~\bibnamefont {Peugeot}},
  \bibinfo {author} {\bibfnamefont {C.}~\bibnamefont {Padurariu}}, \bibinfo
  {author} {\bibfnamefont {C.}~\bibnamefont {Rolland}}, \bibinfo {author}
  {\bibfnamefont {B.}~\bibnamefont {Kubala}}, \bibinfo {author} {\bibfnamefont
  {Y.}~\bibnamefont {Mukharsky}}, \bibinfo {author} {\bibfnamefont
  {Z.}~\bibnamefont {Iftikhar}}, \bibinfo {author} {\bibfnamefont
  {C.}~\bibnamefont {Altimiras}}, \bibinfo {author} {\bibfnamefont
  {P.}~\bibnamefont {Roche}}, \bibinfo {author} {\bibfnamefont
  {H.}~\bibnamefont {le~Sueur}}, \bibinfo {author} {\bibfnamefont
  {P.}~\bibnamefont {Joyez}}, \bibinfo {author} {\bibfnamefont
  {D.}~\bibnamefont {Vion}}, \bibinfo {author} {\bibfnamefont {D.}~\bibnamefont
  {Esteve}}, \bibinfo {author} {\bibfnamefont {J.}~\bibnamefont {Ankerhold}}, \
  and\ \bibinfo {author} {\bibfnamefont {F.}~\bibnamefont {Portier}},\ }\href
  {\doibase 10.1103/PhysRevX.12.021006} {\bibfield  {journal} {\bibinfo
  {journal} {Phys. Rev. X}\ }\textbf {\bibinfo {volume} {12}},\ \bibinfo
  {pages} {021006} (\bibinfo {year} {2022})}\BibitemShut {NoStop}%
\bibitem [{\citenamefont {Hofheinz}\ \emph {et~al.}(2008)\citenamefont
  {Hofheinz}, \citenamefont {Weig}, \citenamefont {Ansmann}, \citenamefont
  {Bialczak}, \citenamefont {Lucero}, \citenamefont {Neeley}, \citenamefont
  {O{'}Connell}, \citenamefont {Wang}, \citenamefont {Martinis},\ and\
  \citenamefont {Cleland}}]{Hofheinz2008Jul}%
  \BibitemOpen
  \bibfield  {author} {\bibinfo {author} {\bibfnamefont {M.}~\bibnamefont
  {Hofheinz}}, \bibinfo {author} {\bibfnamefont {E.~M.}\ \bibnamefont {Weig}},
  \bibinfo {author} {\bibfnamefont {M.}~\bibnamefont {Ansmann}}, \bibinfo
  {author} {\bibfnamefont {R.~C.}\ \bibnamefont {Bialczak}}, \bibinfo {author}
  {\bibfnamefont {E.}~\bibnamefont {Lucero}}, \bibinfo {author} {\bibfnamefont
  {M.}~\bibnamefont {Neeley}}, \bibinfo {author} {\bibfnamefont {A.~D.}\
  \bibnamefont {O{'}Connell}}, \bibinfo {author} {\bibfnamefont
  {H.}~\bibnamefont {Wang}}, \bibinfo {author} {\bibfnamefont {J.~M.}\
  \bibnamefont {Martinis}}, \ and\ \bibinfo {author} {\bibfnamefont {A.~N.}\
  \bibnamefont {Cleland}},\ }\href {\doibase 10.1038/nature07136} {\bibfield
  {journal} {\bibinfo  {journal} {Nature}\ }\textbf {\bibinfo {volume} {454}},\
  \bibinfo {pages} {310} (\bibinfo {year} {2008})}\BibitemShut {NoStop}%
\bibitem [{\citenamefont {Mollow}(1969)}]{Mollow1969Dec}%
  \BibitemOpen
  \bibfield  {author} {\bibinfo {author} {\bibfnamefont {B.~R.}\ \bibnamefont
  {Mollow}},\ }\href {\doibase 10.1103/PhysRev.188.1969} {\bibfield  {journal}
  {\bibinfo  {journal} {Phys. Rev.}\ }\textbf {\bibinfo {volume} {188}},\
  \bibinfo {pages} {1969} (\bibinfo {year} {1969})}\BibitemShut {NoStop}%
\bibitem [{\citenamefont {Schuda}\ \emph {et~al.}(1974)\citenamefont {Schuda},
  \citenamefont {Stroud},\ and\ \citenamefont {Hercher}}]{Schuda1974May}%
  \BibitemOpen
  \bibfield  {author} {\bibinfo {author} {\bibfnamefont {F.}~\bibnamefont
  {Schuda}}, \bibinfo {author} {\bibfnamefont {C.~R.}\ \bibnamefont {Stroud},
  \bibfnamefont {Jr.}}, \ and\ \bibinfo {author} {\bibfnamefont
  {M.}~\bibnamefont {Hercher}},\ }\href {\doibase 10.1088/0022-3700/7/7/002}
  {\bibfield  {journal} {\bibinfo  {journal} {J. Phys. B: At. Mol. Phys.}\
  }\textbf {\bibinfo {volume} {7}},\ \bibinfo {pages} {L198} (\bibinfo {year}
  {1974})}\BibitemShut {NoStop}%
\bibitem [{\citenamefont {Kimble}\ and\ \citenamefont
  {Mandel}(1976)}]{Kimble1976Jun}%
  \BibitemOpen
  \bibfield  {author} {\bibinfo {author} {\bibfnamefont {H.~J.}\ \bibnamefont
  {Kimble}}\ and\ \bibinfo {author} {\bibfnamefont {L.}~\bibnamefont
  {Mandel}},\ }\href {\doibase 10.1103/PhysRevA.13.2123} {\bibfield  {journal}
  {\bibinfo  {journal} {Phys. Rev. A}\ }\textbf {\bibinfo {volume} {13}},\
  \bibinfo {pages} {2123} (\bibinfo {year} {1976})}\BibitemShut {NoStop}%
\bibitem [{\citenamefont {{Ulhaq}}\ \emph {et~al.}(2012)\citenamefont
  {{Ulhaq}}, \citenamefont {{Weiler}}, \citenamefont {{Ulrich}}, \citenamefont
  {{Ro{\ss}bach}}, \citenamefont {{Jetter}},\ and\ \citenamefont
  {{Michler}}}]{Jetter238U}%
  \BibitemOpen
  \bibfield  {author} {\bibinfo {author} {\bibfnamefont {A.}~\bibnamefont
  {{Ulhaq}}}, \bibinfo {author} {\bibfnamefont {S.}~\bibnamefont {{Weiler}}},
  \bibinfo {author} {\bibfnamefont {S.~M.}\ \bibnamefont {{Ulrich}}}, \bibinfo
  {author} {\bibfnamefont {R.}~\bibnamefont {{Ro{\ss}bach}}}, \bibinfo {author}
  {\bibfnamefont {M.}~\bibnamefont {{Jetter}}}, \ and\ \bibinfo {author}
  {\bibfnamefont {P.}~\bibnamefont {{Michler}}},\ }\href {\doibase
  10.1038/nphoton.2012.23} {\bibfield  {journal} {\bibinfo  {journal} {Nat.
  Photonics}\ }\textbf {\bibinfo {volume} {6}},\ \bibinfo {pages} {238}
  (\bibinfo {year} {2012})}\BibitemShut {NoStop}%
\bibitem [{\citenamefont {Gonzalez-Tudela}\ \emph {et~al.}(2013)\citenamefont
  {Gonzalez-Tudela}, \citenamefont {Laussy}, \citenamefont {Tejedor},
  \citenamefont {Hartmann},\ and\ \citenamefont {del
  Valle}}]{Gonzalez_Tudela_2013}%
  \BibitemOpen
  \bibfield  {author} {\bibinfo {author} {\bibfnamefont {A.}~\bibnamefont
  {Gonzalez-Tudela}}, \bibinfo {author} {\bibfnamefont {F.~P.}\ \bibnamefont
  {Laussy}}, \bibinfo {author} {\bibfnamefont {C.}~\bibnamefont {Tejedor}},
  \bibinfo {author} {\bibfnamefont {M.~J.}\ \bibnamefont {Hartmann}}, \ and\
  \bibinfo {author} {\bibfnamefont {E.}~\bibnamefont {del Valle}},\ }\href
  {\doibase 10.1088/1367-2630/15/3/033036} {\bibfield  {journal} {\bibinfo
  {journal} {New J. Phys.}\ }\textbf {\bibinfo {volume} {15}},\ \bibinfo
  {pages} {033036} (\bibinfo {year} {2013})}\BibitemShut {NoStop}%
\bibitem [{\citenamefont {L\'opez Carre\~no}\ \emph {et~al.}(2017)\citenamefont
  {L\'opez Carre\~no}, \citenamefont {del Valle},\ and\ \citenamefont
  {Laussy}}]{article}%
  \BibitemOpen
  \bibfield  {author} {\bibinfo {author} {\bibfnamefont {J.}~\bibnamefont
  {L\'opez Carre\~no}}, \bibinfo {author} {\bibfnamefont {E.}~\bibnamefont {del
  Valle}}, \ and\ \bibinfo {author} {\bibfnamefont {F.}~\bibnamefont
  {Laussy}},\ }\href {\doibase 10.1002/lpor.201700090} {\bibfield  {journal}
  {\bibinfo  {journal} {Laser Photon. Rev.}\ }\textbf {\bibinfo {volume}
  {11}},\ \bibinfo {pages} {1700090} (\bibinfo {year} {2017})}\BibitemShut
  {NoStop}%
\bibitem [{\citenamefont {Sukumar}\ and\ \citenamefont
  {Buck}(1981)}]{SUKUMAR1981211}%
  \BibitemOpen
  \bibfield  {author} {\bibinfo {author} {\bibfnamefont {C.}~\bibnamefont
  {Sukumar}}\ and\ \bibinfo {author} {\bibfnamefont {B.}~\bibnamefont {Buck}},\
  }\href {\doibase https://doi.org/10.1016/0375-9601(81)90825-2} {\bibfield
  {journal} {\bibinfo  {journal} {Phys. Lett. A}\ }\textbf {\bibinfo {volume}
  {83}},\ \bibinfo {pages} {211} (\bibinfo {year} {1981})}\BibitemShut
  {NoStop}%
\bibitem [{\citenamefont {Singh}(1982)}]{PhysRevA.25.3206}%
  \BibitemOpen
  \bibfield  {author} {\bibinfo {author} {\bibfnamefont {S.}~\bibnamefont
  {Singh}},\ }\href {\doibase 10.1103/PhysRevA.25.3206} {\bibfield  {journal}
  {\bibinfo  {journal} {Phys. Rev. A}\ }\textbf {\bibinfo {volume} {25}},\
  \bibinfo {pages} {3206} (\bibinfo {year} {1982})}\BibitemShut {NoStop}%
\bibitem [{\citenamefont {Villas-Boas}\ and\ \citenamefont
  {Rossatto}(2019)}]{Villas-Boas2019Mar}%
  \BibitemOpen
  \bibfield  {author} {\bibinfo {author} {\bibfnamefont {C.~J.}\ \bibnamefont
  {Villas-Boas}}\ and\ \bibinfo {author} {\bibfnamefont {D.~Z.}\ \bibnamefont
  {Rossatto}},\ }\href {\doibase 10.1103/PhysRevLett.122.123604} {\bibfield
  {journal} {\bibinfo  {journal} {Phys. Rev. Lett.}\ }\textbf {\bibinfo
  {volume} {122}},\ \bibinfo {pages} {123604} (\bibinfo {year}
  {2019})}\BibitemShut {NoStop}%
\bibitem [{\citenamefont {Larson}\ and\ \citenamefont
  {Mavrogordatos}(2021)}]{Larson_2021}%
  \BibitemOpen
  \bibfield  {author} {\bibinfo {author} {\bibfnamefont {J.}~\bibnamefont
  {Larson}}\ and\ \bibinfo {author} {\bibfnamefont {T.}~\bibnamefont
  {Mavrogordatos}},\ }\href {\doibase 10.1088/978-0-7503-3447-1} {\emph
  {\bibinfo {title} {The Jaynes{\textendash}Cummings Model and Its
  Descendants}}}\ (\bibinfo  {publisher} {{IOP} Publishing},\ \bibinfo {year}
  {2021})\BibitemShut {NoStop}%
\bibitem [{\citenamefont {Scully}\ and\ \citenamefont
  {Zubairy}(1997)}]{scully_zubairy_1997}%
  \BibitemOpen
  \bibfield  {author} {\bibinfo {author} {\bibfnamefont {M.~O.}\ \bibnamefont
  {Scully}}\ and\ \bibinfo {author} {\bibfnamefont {M.~S.}\ \bibnamefont
  {Zubairy}},\ }\href@noop {} {\emph {\bibinfo {title} {\emph{Quantum
  Optics}}}}\ (\bibinfo  {publisher} {Cambridge University Press},\ \bibinfo
  {address} {Cambridge},\ \bibinfo {year} {1997})\BibitemShut {NoStop}%
\bibitem [{\citenamefont {Vogel}\ and\ \citenamefont
  {Filho}(1995)}]{PhysRevA.52.4214}%
  \BibitemOpen
  \bibfield  {author} {\bibinfo {author} {\bibfnamefont {W.}~\bibnamefont
  {Vogel}}\ and\ \bibinfo {author} {\bibfnamefont {R.~L. d.~M.}\ \bibnamefont
  {Filho}},\ }\href {\doibase 10.1103/PhysRevA.52.4214} {\bibfield  {journal}
  {\bibinfo  {journal} {Phys. Rev. A}\ }\textbf {\bibinfo {volume} {52}},\
  \bibinfo {pages} {4214} (\bibinfo {year} {1995})}\BibitemShut {NoStop}%
\bibitem [{\citenamefont {Gu}\ \emph {et~al.}(2017)\citenamefont {Gu},
  \citenamefont {Kockum}, \citenamefont {Miranowicz}, \citenamefont {xi~Liu},\
  and\ \citenamefont {Nori}}]{GU20171}%
  \BibitemOpen
  \bibfield  {author} {\bibinfo {author} {\bibfnamefont {X.}~\bibnamefont
  {Gu}}, \bibinfo {author} {\bibfnamefont {A.~F.}\ \bibnamefont {Kockum}},
  \bibinfo {author} {\bibfnamefont {A.}~\bibnamefont {Miranowicz}}, \bibinfo
  {author} {\bibfnamefont {Y.}~\bibnamefont {xi~Liu}}, \ and\ \bibinfo {author}
  {\bibfnamefont {F.}~\bibnamefont {Nori}},\ }\href {\doibase
  https://doi.org/10.1016/j.physrep.2017.10.002} {\bibfield  {journal}
  {\bibinfo  {journal} {Physics Reports}\ }\textbf {\bibinfo {volume}
  {718-719}},\ \bibinfo {pages} {1} (\bibinfo {year} {2017})}\BibitemShut
  {NoStop}%
\bibitem [{\citenamefont {Chen}\ \emph {et~al.}(2017)\citenamefont {Chen},
  \citenamefont {Wang}, \citenamefont {Li}, \citenamefont {Tian}, \citenamefont
  {Qiu}, \citenamefont {Inomata}, \citenamefont {Yoshihara}, \citenamefont
  {Han}, \citenamefont {Nori}, \citenamefont {Tsai},\ and\ \citenamefont
  {You}}]{chen2017Singlephotondriven}%
  \BibitemOpen
  \bibfield  {author} {\bibinfo {author} {\bibfnamefont {Z.}~\bibnamefont
  {Chen}}, \bibinfo {author} {\bibfnamefont {Y.}~\bibnamefont {Wang}}, \bibinfo
  {author} {\bibfnamefont {T.}~\bibnamefont {Li}}, \bibinfo {author}
  {\bibfnamefont {L.}~\bibnamefont {Tian}}, \bibinfo {author} {\bibfnamefont
  {Y.}~\bibnamefont {Qiu}}, \bibinfo {author} {\bibfnamefont {K.}~\bibnamefont
  {Inomata}}, \bibinfo {author} {\bibfnamefont {F.}~\bibnamefont {Yoshihara}},
  \bibinfo {author} {\bibfnamefont {S.}~\bibnamefont {Han}}, \bibinfo {author}
  {\bibfnamefont {F.}~\bibnamefont {Nori}}, \bibinfo {author} {\bibfnamefont
  {J.~S.}\ \bibnamefont {Tsai}}, \ and\ \bibinfo {author} {\bibfnamefont
  {J.~Q.}\ \bibnamefont {You}},\ }\href {\doibase 10.1103/PhysRevA.96.012325}
  {\bibfield  {journal} {\bibinfo  {journal} {Phys. Rev. A}\ }\textbf {\bibinfo
  {volume} {96}},\ \bibinfo {pages} {012325} (\bibinfo {year}
  {2017})}\BibitemShut {NoStop}%
\bibitem [{\citenamefont {Brune}\ \emph {et~al.}(1987)\citenamefont {Brune},
  \citenamefont {Raimond}, \citenamefont {Goy}, \citenamefont {Davidovich},\
  and\ \citenamefont {Haroche}}]{PhysRevLett.59.1899}%
  \BibitemOpen
  \bibfield  {author} {\bibinfo {author} {\bibfnamefont {M.}~\bibnamefont
  {Brune}}, \bibinfo {author} {\bibfnamefont {J.~M.}\ \bibnamefont {Raimond}},
  \bibinfo {author} {\bibfnamefont {P.}~\bibnamefont {Goy}}, \bibinfo {author}
  {\bibfnamefont {L.}~\bibnamefont {Davidovich}}, \ and\ \bibinfo {author}
  {\bibfnamefont {S.}~\bibnamefont {Haroche}},\ }\href {\doibase
  10.1103/PhysRevLett.59.1899} {\bibfield  {journal} {\bibinfo  {journal}
  {Phys. Rev. Lett.}\ }\textbf {\bibinfo {volume} {59}},\ \bibinfo {pages}
  {1899} (\bibinfo {year} {1987})}\BibitemShut {NoStop}%
\bibitem [{\citenamefont {Gauthier}\ \emph {et~al.}(1992)\citenamefont
  {Gauthier}, \citenamefont {Wu}, \citenamefont {Morin},\ and\ \citenamefont
  {Mossberg}}]{PhysRevLett.68.464}%
  \BibitemOpen
  \bibfield  {author} {\bibinfo {author} {\bibfnamefont {D.~J.}\ \bibnamefont
  {Gauthier}}, \bibinfo {author} {\bibfnamefont {Q.}~\bibnamefont {Wu}},
  \bibinfo {author} {\bibfnamefont {S.~E.}\ \bibnamefont {Morin}}, \ and\
  \bibinfo {author} {\bibfnamefont {T.~W.}\ \bibnamefont {Mossberg}},\ }\href
  {\doibase 10.1103/PhysRevLett.68.464} {\bibfield  {journal} {\bibinfo
  {journal} {Phys. Rev. Lett.}\ }\textbf {\bibinfo {volume} {68}},\ \bibinfo
  {pages} {464} (\bibinfo {year} {1992})}\BibitemShut {NoStop}%
\bibitem [{\citenamefont {Raimond}\ \emph {et~al.}(2001)\citenamefont
  {Raimond}, \citenamefont {Brune},\ and\ \citenamefont
  {Haroche}}]{RevModPhys.73.565}%
  \BibitemOpen
  \bibfield  {author} {\bibinfo {author} {\bibfnamefont {J.~M.}\ \bibnamefont
  {Raimond}}, \bibinfo {author} {\bibfnamefont {M.}~\bibnamefont {Brune}}, \
  and\ \bibinfo {author} {\bibfnamefont {S.}~\bibnamefont {Haroche}},\ }\href
  {\doibase 10.1103/RevModPhys.73.565} {\bibfield  {journal} {\bibinfo
  {journal} {Rev. Mod. Phys.}\ }\textbf {\bibinfo {volume} {73}},\ \bibinfo
  {pages} {565} (\bibinfo {year} {2001})}\BibitemShut {NoStop}%
\bibitem [{\citenamefont {Blais}\ \emph {et~al.}(2021)\citenamefont {Blais},
  \citenamefont {Grimsmo}, \citenamefont {Girvin},\ and\ \citenamefont
  {Wallraff}}]{RevModPhys.93.025005}%
  \BibitemOpen
  \bibfield  {author} {\bibinfo {author} {\bibfnamefont {A.}~\bibnamefont
  {Blais}}, \bibinfo {author} {\bibfnamefont {A.~L.}\ \bibnamefont {Grimsmo}},
  \bibinfo {author} {\bibfnamefont {S.~M.}\ \bibnamefont {Girvin}}, \ and\
  \bibinfo {author} {\bibfnamefont {A.}~\bibnamefont {Wallraff}},\ }\href
  {\doibase 10.1103/RevModPhys.93.025005} {\bibfield  {journal} {\bibinfo
  {journal} {Rev. Mod. Phys.}\ }\textbf {\bibinfo {volume} {93}},\ \bibinfo
  {pages} {025005} (\bibinfo {year} {2021})}\BibitemShut {NoStop}%
\end{thebibliography}
\providecommand{\noopsort}[1]{}\providecommand{\singleletter}[1]{#1}%

\end{document}